\documentclass[10pt]{article}

\usepackage{epsfig}
\usepackage{amssymb}
\usepackage{amsmath}

\renewcommand{\baselinestretch}{2.0}

\begin{document}

\pagestyle{plain}
\pagenumbering{arabic}
\setcounter{page}{1}

\begin{center}

{\Large \bf Page-Differential Logging: An Efficient and DBMS-independent Approach for Storing Data into Flash Memory}

Yi-Reun Kim$^{*}$, Kyu-Young Whang$^{*}$, Il-Yeol Song$^{**}$ \\
\vspace*{-0.20cm}
$^{*}$Department of Computer Science \\
\vspace*{-0.20cm}
Korea Advanced Institute of Science and Technology\,(KAIST) \\
\vspace*{-0.20cm}
$^{**}$College of Information Science and Technology \\
\vspace*{-0.20cm}
Drexel University \\
\vspace*{-0.20cm}
e-mail:\,$^{*}$\{yrkim, kywhang\}@mozart.kaist.ac.kr, $^{**}$songiy@drexel.edu \\
\end{center}


\renewcommand{\baselinestretch}{1.6}
\begin{abstract}
{\small

Flash memory is widely used as the secondary storage in lightweight
computing devices due to its outstanding advantages over magnetic
disks. Flash memory has many access characteristics different from
those of magnetic disks, and how to take advantage of them is
becoming an important research issue. There are two existing
approaches to storing data into flash memory: page-based and
log-based. The former has good performance for read operations, but
poor performance for write operations. In contrast, the latter has
good performance for write operations when updates are light, but
poor performance for read operations. In this paper, we propose a
new method of storing data, called {\it page-differential logging},
for flash-based storage systems that solves the drawbacks of the two
methods. The primary characteristics of our method are: (1) writing
only the difference\,(which we define as the {\it
page-differential}) between the original page in flash memory and
the up-to-date page in memory; (2) computing and writing the
page-differential only once at the time the page needs to be
reflected into flash memory. The former contrasts with existing
page-based methods that write the whole page including both changed
and unchanged parts of data or from log-based ones that keep track
of the history of all the changes in a page. Our method allows
existing disk-based DBMSs to be reused as flash-based DBMSs just by
modifying the flash memory driver, i.e., it is DBMS-independent.
Experimental results show that the proposed method is superior in
I/O performance, except for some special cases, to existing ones.
Specifically, it improves the performance of various mixes of
read-only and update operations by 0.5\,(the special case when all
transactions are read-only on updated pages) $\sim$ 3.4 times over
the page-based method and by 1.6 $\sim$ 3.1 times over the log-based
one for synthetic data of approximately 1\,Gbytes. The TPC-C
benchmark also shows improvement of the I/O time over existing
methods by 1.2 $\sim$ 6.1 times. This result indicates the
effectiveness of our method under (semi) real workloads.

}
\end{abstract}
\renewcommand{\baselinestretch}{2.0}

\newtheorem{definition}{\bf Definition}
\newtheorem{strategy}{\bf Strategy}
\newtheorem{example}{\bf Example}
\newtheorem{property}{\bf Property}

\newenvironment{newidth}[2]{
 \begin{list}{}{
  \setlength{\topsep}{0pt}
  \setlength{\leftmargin}{#1}
  \setlength{\rightmargin}{#2}
  \setlength{\listparindent}{\parindent}
  \setlength{\itemindent}{\parindent}
  \setlength{\parsep}{\parskip}
 }
\item[]}{\end{list}}

%
%
\section{Introduction}
\label{chap:1}
\vspace*{-0.30cm}
\vspace*{0.30cm} 

Flash memory is a non-volatile secondary storage that is
electrically erasable and reprogrammable\,\cite{CK04,KNM95}. Flash
memory has outstanding advantages over magnetic disks: lighter
weight, smaller size, better shock resistance, lower power
consumption, and faster access time\,\cite{KNM95,LM07,WKC07}. Due to
these advantages, the flash memory is widely used in embedded
systems and mobile devices such as mobile phones, MP3 players, and
digital cameras\,\cite{LM07,NG08}.

Flash memory is much different from a magnetic disk in structures
and access characteristics\,\cite{KV08}. It is composed of a number
of blocks, and each block is composed of a fixed number of pages. It
does not have seek and rotation latency because it is made of
electronic circuits without mechanically moving parts\,\cite{KV08}.
Flash memory provides three kinds of operations\,---\,read, write,
and erase. In order to overwrite existing data in a page, an erase
operation must be performed before writing new data on the
page\,\cite{KV08,LM07}. The write and erase operations are much
slower than the read operation\,\cite{LM07,Sam05}. Besides, the unit
of the erase operation is a block, while the unit of the read and
write operations is a page\,\cite{WKC07}.

There have been a number of
studies\,\cite{YAFFS02,Ban95,KBLLJ06,LCP06,LM07,W01} on the method
of storing updated pages into flash memory for flash-based storage
systems. In this paper, we refer to such methods as {\it page
update} methods. The page update methods are classified into two
categories\,\cite{WKC07}\,---\,page-based\,\cite{Ban95,LCP06} and
log-based\,\cite{YAFFS02,LM07,W01}. Page-based methods write the
whole page into flash memory when an updated page {\it needs to be
reflected} into flash memory\,(e.g., when the page is swapped out
from the DBMS buffer to the database)\,\cite{Ban95,LCP06,WKC07}.
These methods actually read only one page when {\it recreating} a
page from flash memory\,(e.g., reading it into a DBMS buffer). Thus,
they have good read performance. However, they have relatively poor
write performance because they write the whole page including
unchanged parts as well as changed parts of data\,\cite{WKC07}. In
order to overcome this drawback, log-based methods have been
proposed\,\cite{WKC07}. These methods write only the changes\,(which
we call an {\it update log}\,\footnote{An update log contains the
changes in a page resulted in a single update command.}) in the page
into the write buffer, which in turn is written into flash memory
when the buffer is full\,\cite{YAFFS02,LM07,W01}. Thus, compared
with page-based methods, log-based ones have good write performance
when updates are not heavy\,\footnote{\vspace*{-0.2cm} When pages
are frequently updated, the log-based methods could be poorer in
performance as we see in the experiments in p.~30,
Figure~\ref{fig:5_experiment2}.}\,\cite{WKC07}. Log-based methods,
however, have relatively poor read performance because they keep the
history of all the changes\,(i.e., multiple update logs) in a page.
Whenever an update is done, they write an update log into the write
buffer. Thus, when updates are done multiple times, the update logs
are likely to be written into multiple pages in flash memory. Thus,
log-based methods need to read multiple pages when recreating a page
from flash memory.

In this paper, we propose a page update method called {\it
page-differential logging}\,({\it PDL}) for flash-based storage
systems. A {\it page-differential}\,(simply, a {\it differential})
is defined as the difference between the original page in the flash
memory and the up-to-date page in memory. This novel method is much
different from page-based methods or log-based ones in the following
ways. (1) We write only the differential of an updated page. This
characteristic stands in contrast with page-based methods that write
the whole page including changed and unchanged parts of data or
log-based ones that keep track of the history of all the
changes\,(i.e., multiple update logs) in a page. Furthermore, we
compute and write the differential only once at the time the updated
page needs to be reflected into flash memory. The overhead of
generating the differential is relatively minor because, in flash
memory, the speed of read operation is much faster than those of
write or erase operations. (2) When recreating a page from flash
memory, we need fewer read operations than log-based ones do because
we read at most two pages: the original page and the single page
containing the differential. (3) When we need to reflect an updated
page into flash memory, we need fewer write operations than others
do because we write only the differential. A side benefit is that
the longevity of flash memory is also improved due to fewer erase
operations resulted from fewer write operations. (4) Our method is
loosely-coupled with the storage system while the log-based ones are
tightly-coupled. The log-based methods need to modify the storage
management module of the DBMS because they must identify the changes
in a page whenever it is updated. These changes can be identified
only inside the storage management module because they are
internally maintained by the system. On the other hand, our method
does not need to modify the module of the DBMS because it computes
the differential outside the storage management module by comparing
the page that needs to be reflected with the original page in the
flash memory. We elaborate on this point later in
Section~\ref{chap:4}.

The contributions of this paper are as follows. (1) we propose a new
notion of ``differential'' of a page. Using this notion, we then
propose a new approach to updating pages that we call {\it
page-differential logging}. (2) Our method is DBMS-independent.
(3) Through extensive experiments, we show that the overall read and
write performance of our method is mostly superior to those of
existing ones.

Hereafter, in order to reduce ambiguity in this paper, we
distinguish logical pages from physical pages. We call the pages in
memory {\it logical pages} and the ones in flash memory {\it
physical pages}. For ease of exposition, we assume that the size of
a logical page is equal to that of a physical page.

The rest of this paper is organized as follows. Section~\ref{chap:2}
introduces flash memory. Section~\ref{chap:3} describes prior work
related to the page update methods for flash-based storage systems.
Section~\ref{chap:4} presents a new page update method called {\it
page-differential logging}. Section~\ref{chap:5} presents the
results of performance evaluation. Section~\ref{chap:6} summarizes
and concludes the paper.

%
%
\section{Flash Memory}
\label{chap:2}
\vspace*{-0.30cm}

Based on the structure of memory cells, there are two major types of
flash memory\,\cite{GT05a}: the NAND type and the NOR type. The
former is suitable for storing data, and the latter for storing
code\,\cite{PSSKK03}. In the rest of this paper, we use the term
`flash memory' to indicate the NAND type flash memory, which is
widely used in flash-based storage
systems\,\footnote{\vspace*{-0.2cm} In this paper, we focus on flash
memory but not on solid state disks\,({\it SSD's})\,\cite{Sam08},
which have controllers with their own page update methods.}.

\begin{figure}[h!]
  \vspace*{0.50cm}
  \centerline{\psfig{file=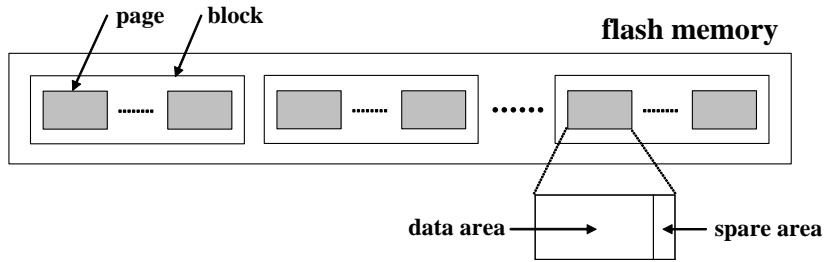, width=11cm}}
  \vspace*{-0.2cm}
  \caption{The structure of flash memory.}
  \label{fig:2_flashmemory}
\end{figure}

Figure~\ref{fig:2_flashmemory} shows the structure of flash memory.
The flash memory consists of $N_{block}$ {\it blocks}, and each
block consists of $N_{page}$ {\it pages}. A page is the smallest
unit of reading and writing data, and a block is the smallest unit
of erasing data\,\cite{WKC07}. Each page consists of a data area
used for storing data and a spare area used for storing auxiliary
information such as the valid bit, obsolete bit, bad block
identification, and error correction check\,(ECC)\,\cite{PSSKK03}.

\clearpage 

We consider three operations: read, write, and erase\,\cite{GT05a}.

\vspace*{-0.3cm}
\begin{itemize}
\item \textbf{The read operation\,:} returns all the bits in the addressed page

\item \textbf{The write operation\,:} changes a set of bits selected in the target page from 1 to 0

\item \textbf{The erase operation\,:} sets all the bits in the addressed block to 1

\vspace*{-0.3cm}
\end{itemize}

\noindent The operations in flash memory are different from those in
the magnetic disk in two ways. First, all the bits in flash memory
are initially set to 1. Thus, writing to flash memory means
selectively changing some bits in a page from 1 to 0. Next, the
erase operation in flash memory changes the bits in a block back to
1. Each block can sustain only a limited number of erase operations
before becoming unreliable, which is restricted to about
100,000\,\footnote{\vspace*{-0.2cm} Due to this characteristic,
there have been a number of studies on wear-leveling\,\cite{KNM95}
and bad block management\,\cite{PSSKK03}. \vspace*{-0.2cm} However,
we do not address them in this paper, but these studies can be
applied to the storage system independently of the page update
methods discussed in this paper.}\,\cite{LM07,NG08}.

Due to the restriction of the write and erase operations, a write
operation is usually preceded by an erase operation in order to
overwrite a page\,\cite{KV08,LM07}. We first change all the bits in
the block to 1 using an erase operation, and then, change some bits
in the page to 0 using a write operation. We note that the erase
operation is performed in a much larger unit than a write operation,
i.e., the former is performed on a block while the latter on a page.
The specific techniques for overwriting a page depend on the page
update method employed. These techniques are discussed in
Section~\ref{chap:3}.

Based on the capacity of memory cells, there are two types of flash
memory\,\cite{KV08}: Single Level Cell\,(SLC)-type and Multi Level
Cell\,(MLC)-type. The former is capable of storing one data bit per
cell, while the latter is capable of storing two\,(or even more)
data bits per cell. Thus, MLC-type flash memory has greater capacity
than SLC-type one and is expected to be widely used in high-capacity
flash storages\,\cite{KV08}. Table~\ref{tbl:2_flashmemory}
summarizes the parameters and values of MLC flash memory we use in
our experiments. We note that the size of a page is 2,048 bytes, and
a block has 64 pages. In addition, the access time of operations
increases in the following order: read, write, and erase. The read
operation is 9.2 times faster than the write operation, which is 1.5
times faster than the erase operation.

\vspace*{0.50cm}
\renewcommand{\baselinestretch}{1.10}
\begin{table}
\begin{center}
\caption{The parameters and values of flash memory$^{*}$.}
\vspace*{0.3cm}
\begin{tabular} {|c|l|c|}
\hline
\multicolumn{1}{|c|}{Symbols} & \multicolumn{1}{c|}{Definitions} & \multicolumn{1}{c|}{Values}\\
\hline \hline
$ N_{block} $ & the number of blocks & $32,768$ \\
\hline
$ N_{page}  $ & the number of pages in a block & $64$ \\
\hline
$ S_{block} $ & the size of a block (bytes)~($=\,N_{page}\,\times\,S_{page}$) & $135,168$ ($64\,\times\,2,112$)\\
\hline
$ S_{page}  $ & the size of a page (bytes)~($=\,S_{data}\,+\,S_{spare}$) & $2,112~(=\,2,048\,+\,64)$ \\
\hline
$ S_{data}  $ & the size of data area in a page (bytes) & $2,048$ \\
\hline
$ S_{spare} $ & the size of spare area in a page (bytes) & $64$ \\
\hline
$ T_{read}  $ & the read time for a page ($\mu s$) & $110$ \\
\hline
$ T_{write} $ & the write time for a page ($\mu s$) & $1010$ \\
\hline
$ T_{erase} $ & the erase time for a block ($\mu s$) & $1500$ \\
\hline
\end{tabular}
\label{tbl:2_flashmemory}
\end{center}
\hspace*{1.5cm} $^{*}$ Samsung K9L8G08U0M 2\,Gbytes MLC NAND flash memory\,\cite{Sam05}
\end{table}
\renewcommand{\baselinestretch}{2.0}

%
%
\section{Related Work}
\label{chap:3}
\vspace*{-0.30cm}



\noindent \textbf{\large The Page-Based Approach}

\begin{figure}[h!]
  \vspace*{0.50cm}
  \centerline{\psfig{file=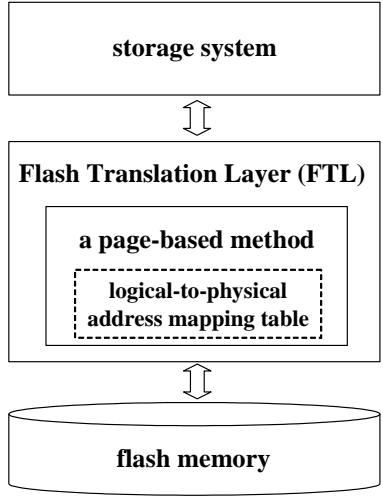, width=5cm}}
  \vspace*{-0.2cm}
  \caption{The architecture of the page-based method.}
  \label{fig:3_ftl}
\end{figure}

In page-based methods\,\cite{Ban95,LCP06}, a logical page is stored
into a physical page. When an updated logical page needs to be
reflected into flash memory, the whole logical page is written into
a physical page\,\cite{WKC07}. When a logical page is recreated from
flash memory, it is read directly from a physical page. These
methods are loosely-coupled with the storage system because they can
be implemented in a middle layer, called the {\it Flash Translation
Layer}\,({\it FTL})\,\cite{Ban95}, which maintains
logical-to-physical address mapping between logical and physical
pages as shown in Figure~\ref{fig:3_ftl}. The FTL can be implemented
as hardware in the controller residing in SSD's, or can be
implemented as software in the operating system for embedded
boards\,\footnote{Commercial FTL's for SSD's or embedded boards
typically use page-based methods\,\cite{APWDMP08}}.

In page-based methods, there are two update
schemes\,\cite{NG08}\,---\,in-place update and out-place
update\,---\,depending on whether or not the logical page is always
written into the same physical page. When a logical page needs to be
reflected into flash memory, the in-place update overwrites it into
the specific physical page that was read\,\cite{NG08}, but the
out-place update writes it into a new physical
page\,\cite{CK04,WKC07}.

~\newline \noindent \textbf{In-Place Update:} As explained in
Section~\ref{chap:2}, the write operation in flash memory cannot
change bits in a page to 1. Therefore, when overwriting the logical
page $l_{1}$ that was read from the physical page $p_{1}$ in the
block $b_{1}$ into the same physical page $p_{1}$, we do the
following four steps: (1) read all the pages in $b_{1}$ except
$p_{1}$; (2) erase $b_{1}$; (3) write $l_{1}$ into $p_{1}$; (4)
write all the pages read in Step\,(1) except $l_{1}$ in the
corresponding pages in $b_{1}$. The in-place update scheme suffers
from severe performance problems and is rarely used in flash
memory\,\cite{NG08} because it causes an erase operation and
multiple read and write operations whenever we need to reflect a
logical page into flash memory.

\begin{figure}[h!]
  \vspace*{0.50cm}
  \centerline{\psfig{file=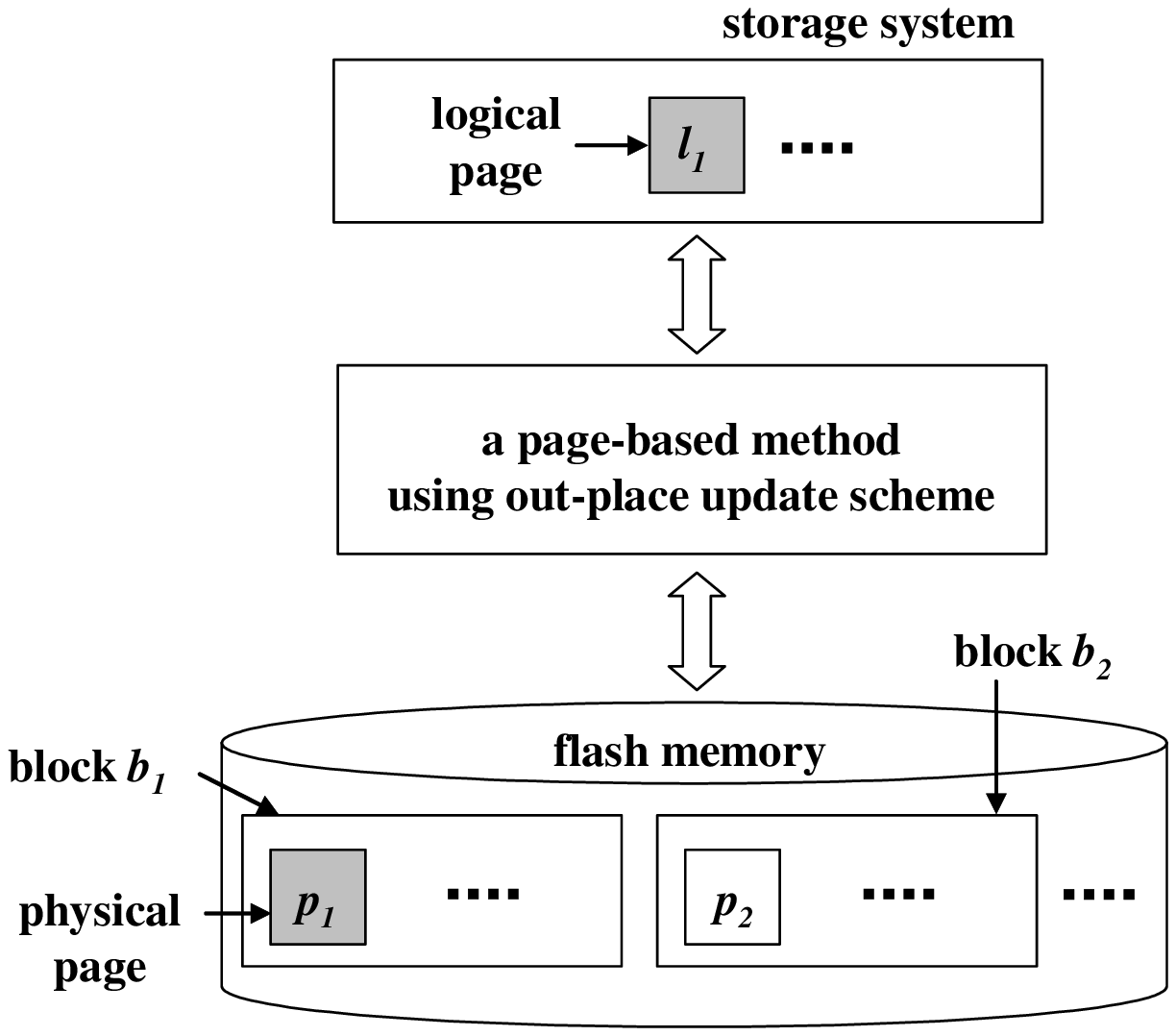, width=7.5cm}
              \psfig{file=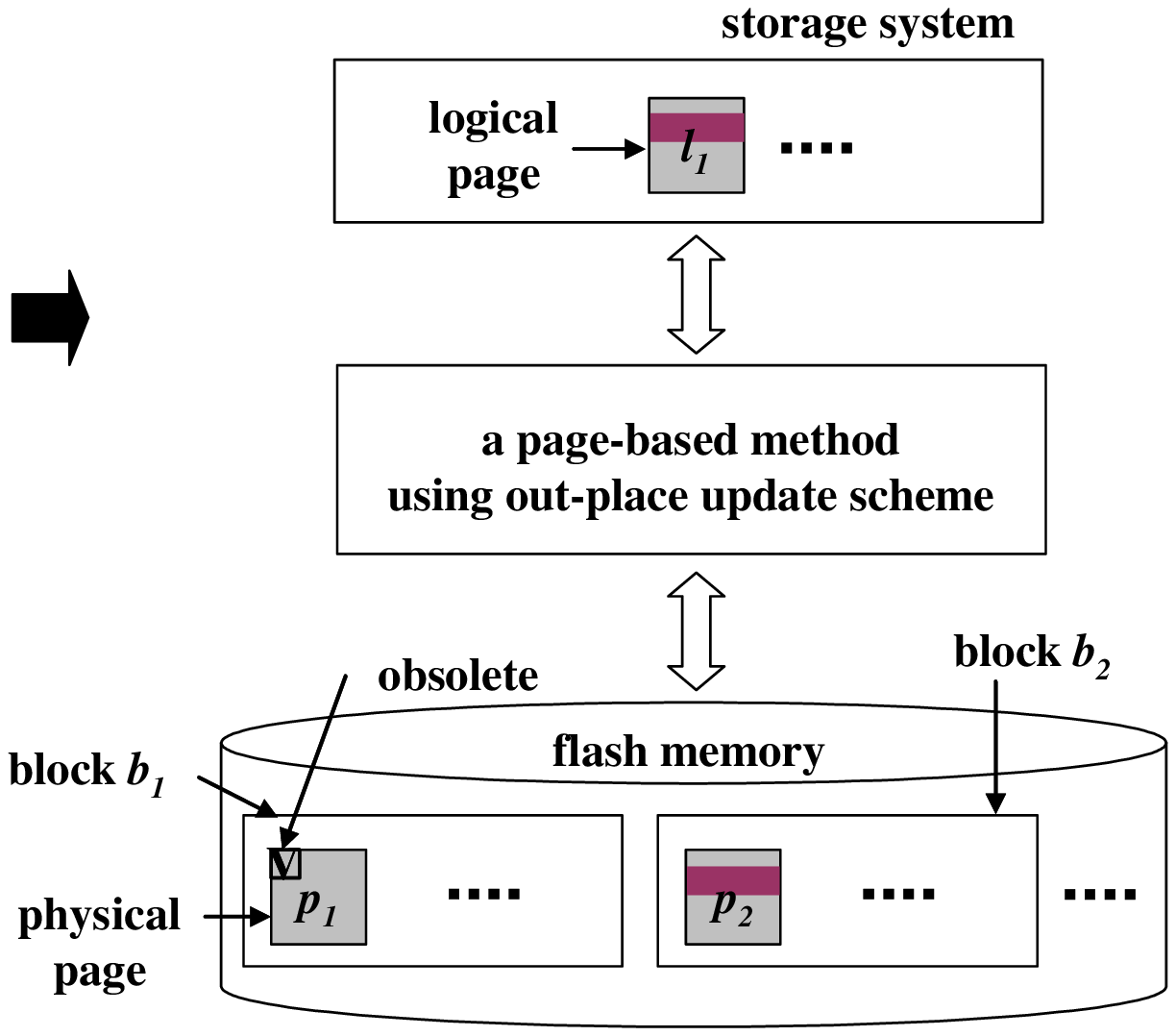, width=7.5cm}}
  \centerline{\hspace*{1.5cm} (a) The logical page $l_{1}$ read from \hspace*{2.0cm}
              (b) The updated logical page $l_{1}$ and \hspace*{0.0cm}}
  \vspace*{-0.2cm}
  \centerline{\hspace*{3.0cm} the physical page $p_{1}$. \hspace*{3.0cm}
              the process of writing it into the physical page $p_{2}$. \hspace*{0.0cm}}
  \vspace*{-0.2cm}
  \caption{An example of out-place update.}
\label{fig:3_outplace}
\end{figure}

~\newline \noindent \textbf{Out-Place Update:}
Figure~\ref{fig:3_outplace} shows a typical example of the out-place
update scheme. Figure~\ref{fig:3_outplace}\,(a) shows the logical
page $l_{1}$ read from the physical page $p_{1}$ in the block
$b_{1}$. Figure~\ref{fig:3_outplace}\,(b) shows the updated logical
page $l_{1}$ and the two physical pages $p_{1}$ and
$p_{2}$\,---\,the original page read and the new page written. In
order to overcome the drawback of in-place update, when we need to
reflect the logical page $l_{1}$ into flash memory, the out-place
update scheme first writes $l_{1}$ into a new physical page $p_{2}$,
and then, sets $p_{1}$ to obsolete\,\footnote{We set a page to
obsolete by changing the obsolete bit in the spare area of the page
from 1 to 0 as in Gal et al.\,\cite{GT05a}.}. When there is no more
free page in flash memory, a block is selected and obsolete pages in
it are reclaimed by garbage collection\,\cite{GT05a}, which converts
obsolete pages to free pages. The out-place update scheme is widely
used in flash-based storage systems\,\cite{WKC07} because it does
not cause an erase operation when a logical page is to be reflected
into flash memory.

~\newline \noindent \textbf{\large The Log-Based Approach}

In log-based methods\,\cite{YAFFS02,LM07,W01}, a logical page is
generally stored into multiple physical pages\,\cite{LM07}. Whenever
logical pages are updated, the update logs of multiple logical pages
are first collected into a write buffer in memory\,\cite{WKC07}.
When this buffer is full, it is written into a single physical page.
Thus, when a logical page is updated many times, its update logs can
be stored into multiple physical pages. Accordingly, when recreating
a single logical page, multiple physical pages may need to be read
and merged.
The log-based methods are tightly-coupled with the storage system
because the storage system must be modified to be able to identify
the update logs of a logical page.

Among log-based methods, there are {\it Log-structured File
system}\,({\it LFS})\,\cite{RO92}, {\it Journaling Flash File
System}\,({\it JFFS})\,\cite{W01}, {\it Yet Another Flash File
System}\,({\it YAFFS})\,\cite{YAFFS02}, and {\it In-Page
Logging}\,({\it IPL})\,\cite{LM07}. In LFS, JFFS, and YAFFS, the
update logs of a logical page can be written into arbitrary log
pages in flash memory while, in IPL, the update logs should be
written into specific log pages. IPL divides the pages in each block
into a fixed number of {\it original pages} and {\it log pages}. It
writes the update logs of a logical page into only the log pages in
the block containing the original (physical) page of the logical
page. Therefore, when recreating the logical page, IPL reads the
original page and only the log pages in the same block. When there
is no free log page in the block, IPL merges the original pages with
the log pages in the block, and then, writes the merged pages into
pages in a new block\,(this process is called {\it
merging}\,\cite{LM07}). The old block is subsequently erased and
garbage-collected. Consequently, IPL improves read performance by
reducing the number of log pages to read from flash memory when
recreating a logical page because log pages do not increase
indefinitely\,(i.e., is bound) due to merging. The performance of
IPL is similar to other log-based methods since IPL inherits the
advantages and drawbacks of log-based methods other than the effect
of merging and bound read performance.

\begin{figure}[h!]
  \centerline{\psfig{file=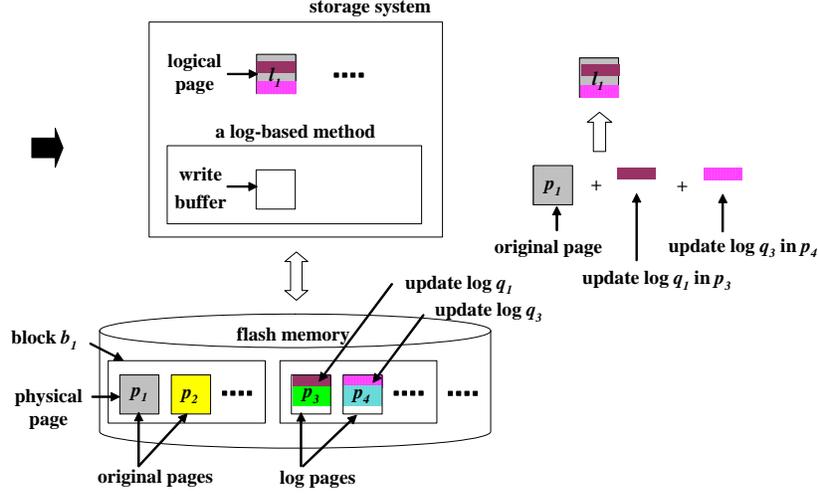, width=11cm}}
  \centerline{(d) The logical page $l_{1}$ being recreated from flash memory.}
  \vspace*{-0.2cm}
  \caption{An example of the log-based approach.}
  \label{fig:3_logbased}
\end{figure}

Figure~\ref{fig:3_logbased} shows a typical example of the log-based
methods. Figure~\ref{fig:3_logbased}\,(a) shows the logical pages
$l_{1}$ and $l_{2}$ in memory. Figure~\ref{fig:3_logbased}\,(b)
shows the update logs $q_{1}$ and $q_{2}$ of logical pages $l_{1}$
and $l_{2}$, respectively, and the process of writing them into
flash memory. Here, the update logs $q_{1}$ and $q_{2}$ are first
written into the write buffer, and then, the content of the write
buffer is written into the log page $p_{3}$. Thus, the update logs
$q_{1}$ and $q_{2}$ are collected into the same log page $p_{3}$.
Figure~\ref{fig:3_logbased}\,(c) shows a similar situation for the
update logs $q_{3}$ and $q_{4}$ of logical pages $l_{1}$ and
$l_{2}$. Figure~\ref{fig:3_logbased}\,(d) shows the logical page
$l_{1}$ being recreated from flash memory. Here, $l_{1}$ is
recreated by merging the original page $p_{1}$ with the update logs
$q_{1}$ and $q_{3}$ read from the log pages $p_{3}$ and $p_{4}$,
respectively.

\begin{figure}[h!]
  \vspace*{0.50cm}
  \centerline{\psfig{file=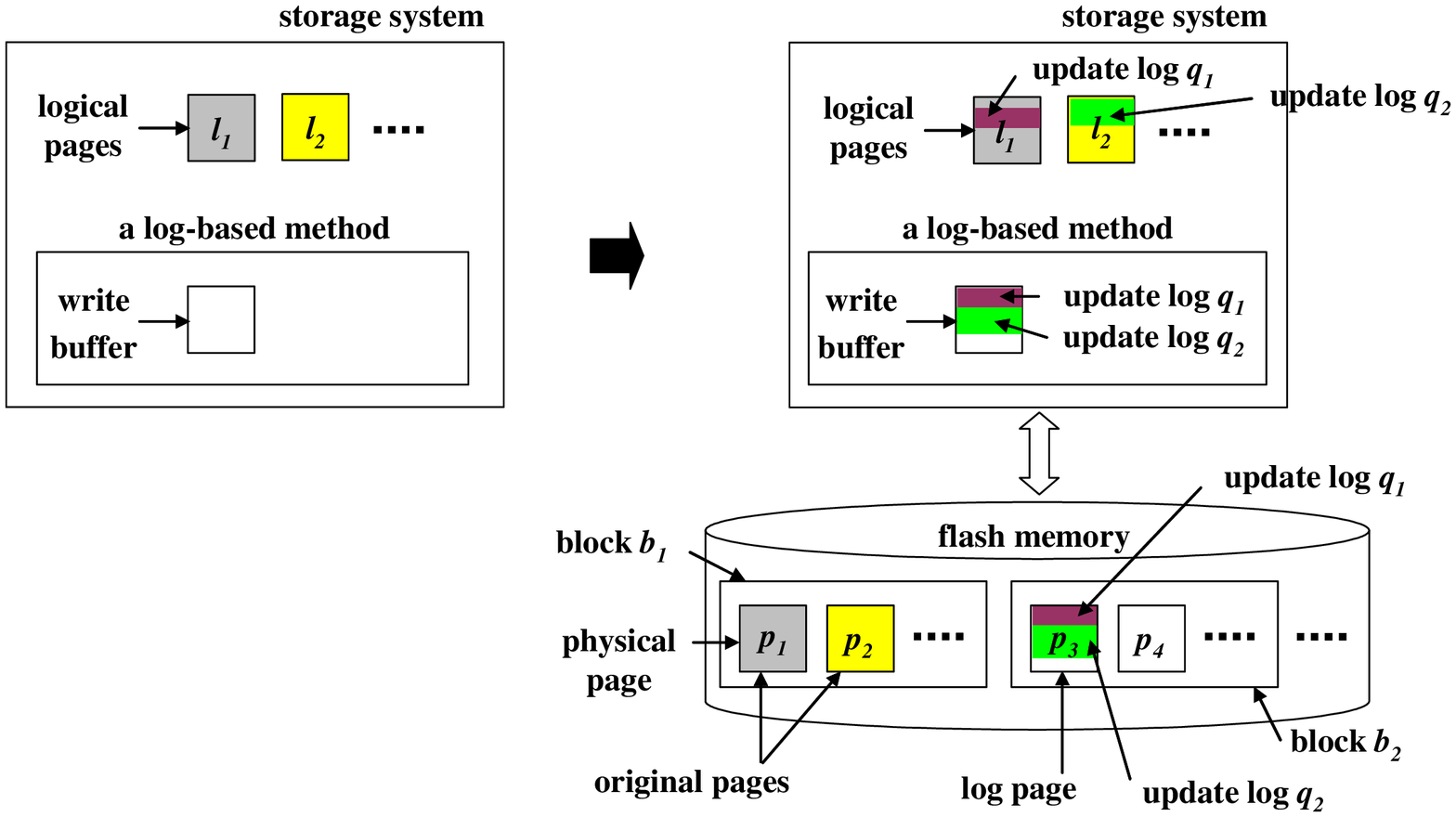, width=11.5cm}}
  \centerline{\hspace*{0.0cm} (a) The logical pages $l_{1}$ and $l_{2}$ \hspace*{1.5cm}
              (b) The update logs $q_{1}$ and $q_{2}$ of logical pages $l_{1}$ and $l_{2}$, and}
  \vspace*{-0.2cm}
  \centerline{\hspace*{2.0cm} in memory. \hspace*{2.75cm}
              the process of writing them into the log page $p_{3}$ in flash memory. \hspace*{0.0cm}}
  \vspace*{0.5cm}
  \centerline{\psfig{file=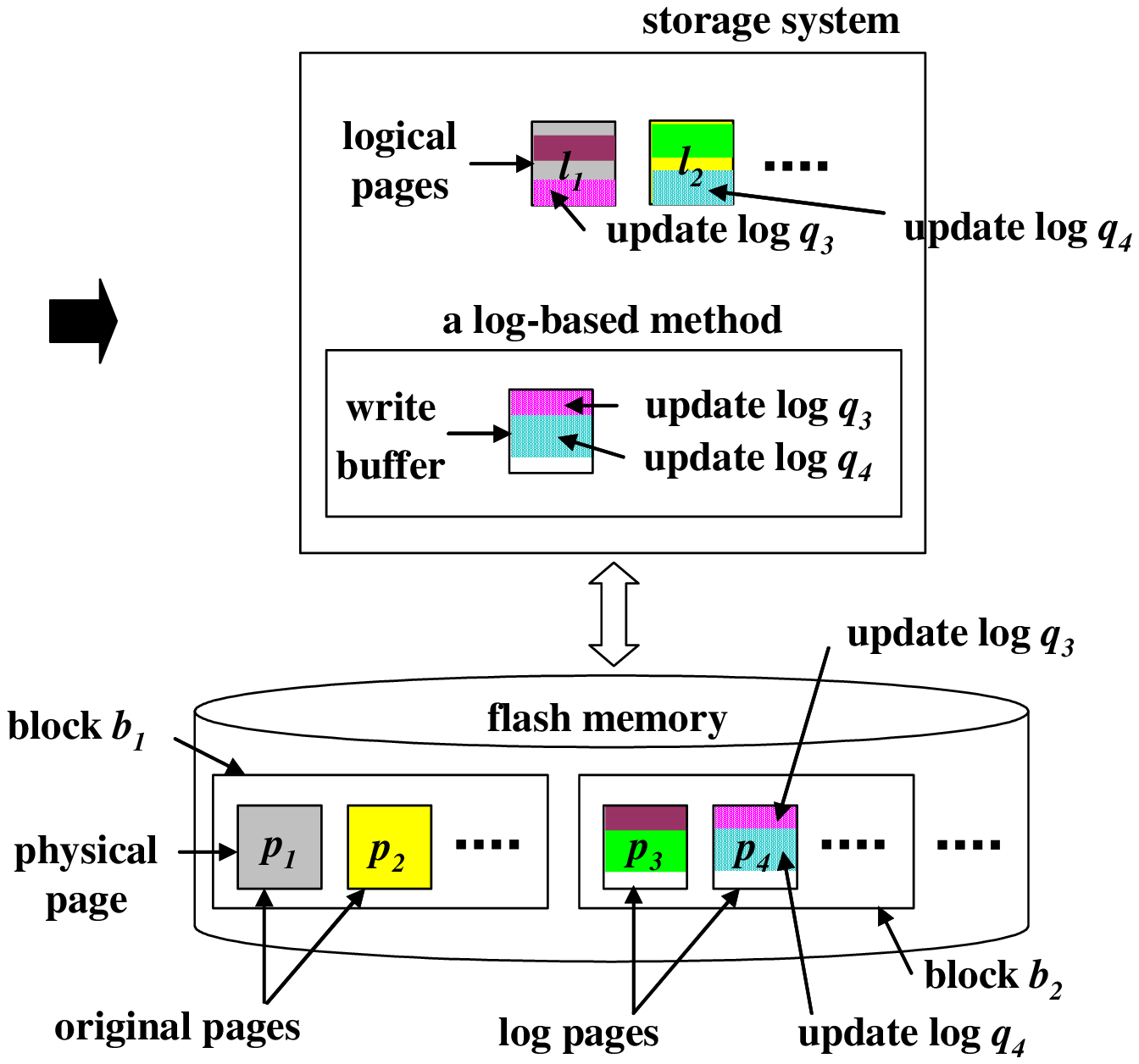, width=7.5cm}}
  \centerline{(c) The update logs $q_{3}$ and $q_{4}$ of logical pages $l_{1}$ and $l_{2}$,}
  \vspace*{-0.2cm}
  \centerline{and the process of writing them into the log page $p_{4}$ in flash memory.}
  \vspace*{4.0cm} 
\end{figure}

\section{The Page-Differential Logging Approach}
\label{chap:4}
\vspace*{-0.30cm}

In this section, we propose {\it page-differential logging}\,({\it
PDL}) for flash-based storage systems. Section~\ref{chap:4_1}
explains the design principles, and then, presents PDL, which
conforms to these principles. Section~\ref{chap:4_2}
and~\ref{chap:4_3} present the data structures and algorithms.
Section~\ref{chap:4_4} discusses the strengths and limitations.

\vspace*{-0.1cm}
\subsection{Design Principles}
\label{chap:4_1}
\vspace*{-0.1cm}

We identify three design principles for PDL in order to guarantee
good performance for both read and write operations. These
principles overcome the drawbacks of both the page-based methods and
the log-based methods in the following ways.

\vspace*{-0.3cm}
\begin{itemize}
\item \textbf{writing difference only\,:} We write only the {\it difference} when a logical page needs to be reflected into flash memory.

\item \textbf{at-most-one-page writing\,:} We write {\it at most one} physical page when a logical page needs to be reflected into flash memory even if the page has been updated in memory multiple times.

\item \textbf{at-most-two-page reading\,:} We read {\it at most two} physical pages when recreating a logical page from flash memory.
\end{itemize}

Page-differential logging method conforms to these three design
principles. In this method, a logical page is stored into two
physical pages\,---\,a {\it base page} and a {\it differential
page}. Here, the base page contains a whole logical page, which
could be the old version, and the differential page contains the
difference between the base page and the up-to-date logical page.
A differential page can contain differentials of multiple logical
pages. Thus, the differentials of two logical pages could be stored
in the same differential page.

\vspace*{-0.15cm} 

The differential has the following advantages over the list of
update logs in the log-based methods. (1) It can be computed without
maintaining all the update logs, i.e., it can be computed by
comparing the updated logical page with its base page only when the
updated logical page needs to be reflected into flash memory. (2) It
contains only the difference from the original page for the part
that has been updated multiple times in a logical page. When a
specific part in a logical page is updated in memory multiple times,
the list of update logs contains all the history of changes while
the differential contains only the difference between original data
and the up-to-date data. For instance, let us assume that a logical
page is updated in memory twice as follows: $...\,aaaaaa\,...
\rightarrow ...\,bbbbba\,... \rightarrow ...\,bcccba\,...$. Here,
the list of update logs contains two changes $bbbbb$ and $ccc$ while
the differential contains only the difference $bcccb$.

\vspace*{-0.15cm} 

In PDL, when an updated logical page needs to be reflected into
flash memory, we create a differential by comparing the logical page
with the base page in flash memory, and then, write the differential
into the one-page write buffer, which is subsequently written into
flash memory when it is full. Therefore, it conforms to the
\textbf{writing-difference-only} principle.

\vspace*{-0.15cm} 

We note that, when a logical page is simply updated, we just update
the logical page in memory without recording the log. Instead, we
defer creating and writing the differential until the updated
logical page needs to be reflected into flash memory. Thus, our
method satisfies the \textbf{at-most-one-page writing} principle.

Theoretically, the size of the differential cannot be larger than
that of one page. However, practically, it could be larger if a
large part of the page has been updated. This case can occur since
the differential contains not only the changed data but also the
meta data such as offsets and lengths. In this case, we discard the
created differential and write the updated logical page itself into
flash memory as a new base page in order to satisfy the
\textbf{at-most-one-page writing} principle. (In this special case,
PDL becomes the same as the page-based method.)

When recreating a logical page from flash memory, we read the base
page and its corresponding differential page, and then, merge the
base page with its differential in the differential page. However,
we need to read only one physical page if the base page has not been
updated\,(i.e., there is no differential page). Thus, we need to
read at most two physical pages, and accordingly, PDL conforms to
the \textbf{at-most-two-page reading} principle.

When there is no more free page in flash memory, obsolete pages are
reclaimed by garbage collection. Here, we select one block for
garbage collection. Since it may contain valid base or differential
pages, before erasing the block, we move those valid pages into a
new block, which is reserved for the garbage collection
process\,\cite{GT05a}. For differential pages, however, we move only
valid differentials into a new differential page, i.e., we do
compaction here. Our method requires fewer write operations than
page-based or log-based ones do because it satisfies the
writing-difference-only and at-most-one-page writing principles.
Thus, our method invokes garbage collection less frequently than
other methods do.

\begin{figure}[h!]
  \vspace*{0.50cm}
  \centerline{\psfig{file=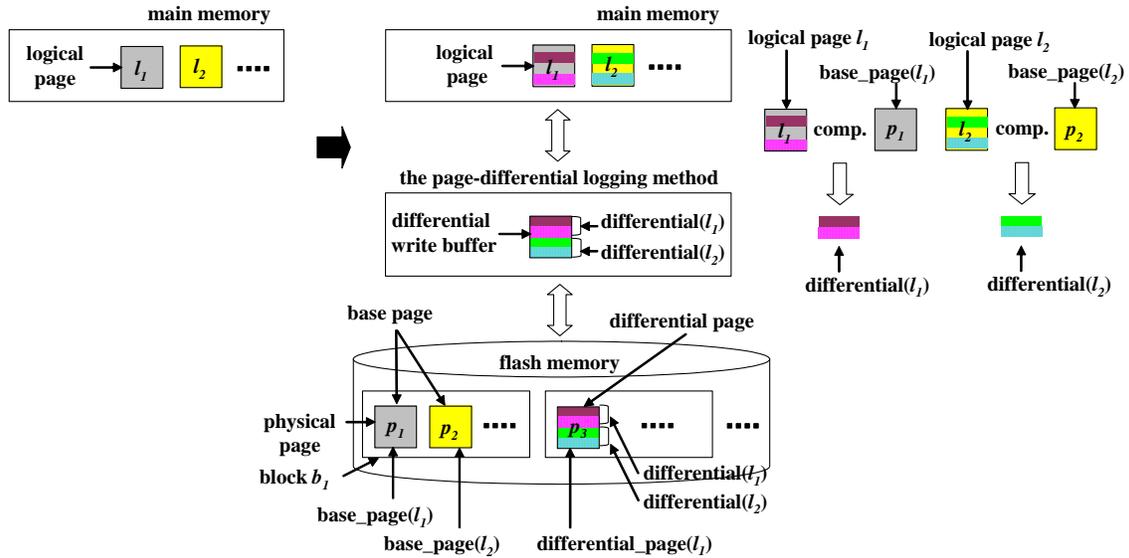, width=15cm}}
  \centerline{\hspace*{0.5cm} (a) The logical pages $l_{1}$ and $l_{2}$ \hspace*{1.5cm}
              (b) The updated logical pages $l_{1}$ and $l_{2}$, and the process of\hspace*{2.0cm}}
  \vspace*{-0.2cm}
  \centerline{\hspace*{0.0cm} in memory. \hspace*{3.5cm}
              writing them into the differential page $p_{3}$ in flash memory. \hspace*{0.0cm}}
  \vspace*{0.5cm}
  \centerline{\psfig{file=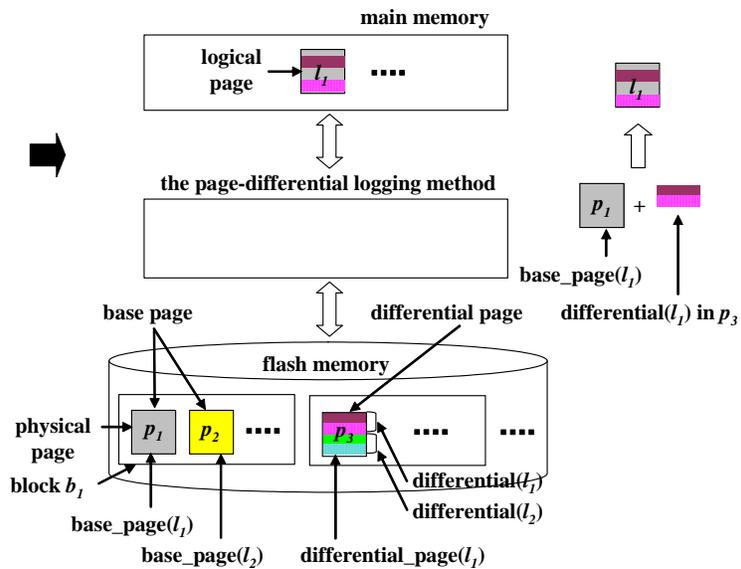, width=10cm}}
  \centerline{(c) The logical page $l_{1}$ recreated from flash memory.}
  \vspace*{-0.2cm}
  \caption{An example of the differential-based approach.}
\label{fig:4_differentiallogging}
\end{figure}

Figure~\ref{fig:4_differentiallogging} shows an example of PDL.
Here, we have base\_page({\it p}), differential\_page({\it p}), and
differential({\it p}) for the logical page {\it p}.
Figure~\ref{fig:4_differentiallogging}\,(a) shows the logical pages
$l_{1}$ and $l_{2}$ in memory.
Figure~\ref{fig:4_differentiallogging}\,(b) shows the updated
logical pages $l_{1}$ and $l_{2}$, and the process of writing them
into flash memory. When $l_{1}$ and $l_{2}$ need to be reflected
into flash memory, we perform the following three steps: (1) read
the base pages $p_{1}$ and $p_{2}$ from flash memory; (2) create
differential($l_{1}$) and differential($l_{2}$) by comparing $l_{1}$
and $l_{2}$ with the base pages $p_{1}$ and $p_{2}$, respectively;
(3) write differential($l_{1}$) and differential($l_{2}$) into the
write buffer, which is subsequently written into the physical page
$p_{3}$ when the buffer is full. We note that $l_{1}$ and $l_{2}$
from different logical pages are written into the same differential
page $p_{3}$. Figure~\ref{fig:4_differentiallogging}\,(c) shows the
logical page $l_{1}$ recreated from flash memory by merging the base
page $p_{1}$ with differential($l_{1}$) in
$p_{3}$\,\footnote{\vspace*{-0.2cm} Conceptually, we require an
assembly buffer in order to merge the base page with the
differential. But, in practice, we can use the logical page itself
as the assembly buffer.}.

\vspace*{-0.1cm}
\subsection{Data Structures}
\label{chap:4_2}
\vspace*{-0.1cm}

The data structures used in {\it flash memory} are base pages,
differential pages, and differentials. A base page stores a logical
page in its data area and stores the page's type, physical page ID,
and creation time stamp in its spare area. Here, the {\it type}
indicates whether the page is a base one or differential one, and
the physical page ID represents the unique identifier of a page in
the database. The {\it creation time stamp} indicates when the base
page was created.

A differential page stores differentials of logical pages in its
data area and stores the page's type in its spare area. A physical
page ID and a creation time stamp are stored also in a differential
to identify the base page to which the differential belongs and when
the differential was created. Therefore, the structure of a
differential is in the form of $<$\,{\it physical page ID,} {\it
creation time stamp, }[\,{\it offset, length, changed
data}\,]$^{+}$$>$.

The three data structures used in {\it memory} are the {\it physical
page mapping table}, the {\it valid differential count table}, and
the {\it differential write buffer}. The {\it physical page mapping
table} maps a physical page ID into $<$\,{\it base page address,
differential page address}\,$>$. This table is used to indirectly
reference a base and differential page pair in flash memory because,
in flash memory, the positions of the physical pages can be changed
by the out-place scheme.

The {\it valid differential count table} counts the number of valid
differentials\,(i.e., those that have not been obsoleted) in a
differential page. When the count becomes 0, the differential page
is set to obsolete and made available for garbage collection.

\begin{figure}[h!]
  \vspace*{0.50cm}
  \centerline{\psfig{file=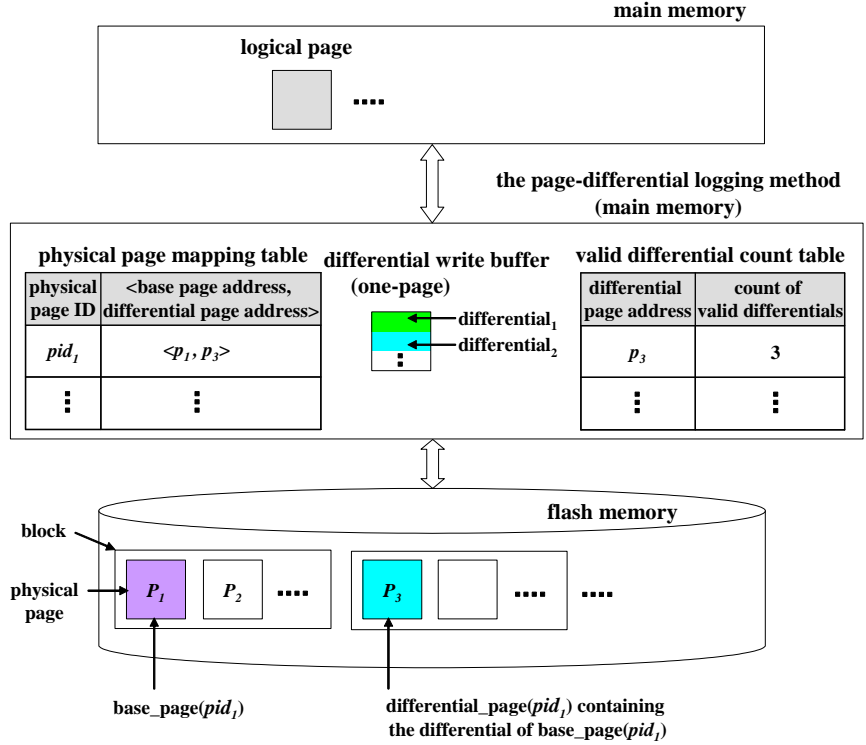, width=11.5cm}}
  \vspace*{-0.2cm}
  \caption{The data structures for PDL.}
  \label{fig:4_differentiallogging_datastructure}
\end{figure}

The {\it differential write buffer} is used to collect differentials
of logical pages into memory and later write them into a
differential page in flash memory when it is full. The differential
write buffer consists of a single page, and thus, the memory usage
is negligible. Figure~\ref{fig:4_differentiallogging_datastructure}
shows the data structures for PDL.

\vspace*{-0.1cm}
\subsection{Algorithms}
\label{chap:4_3}
\vspace*{-0.1cm}

In this section, we present the algorithms for writing a logical
page into flash memory and for recreating a logical page from flash
memory. We call them {\it PDL\_Writing} and {\it PDL\_Reading},
respectively.

\begin{figure}[h!]
  \vspace*{0.50cm}
  \centerline{\psfig{file=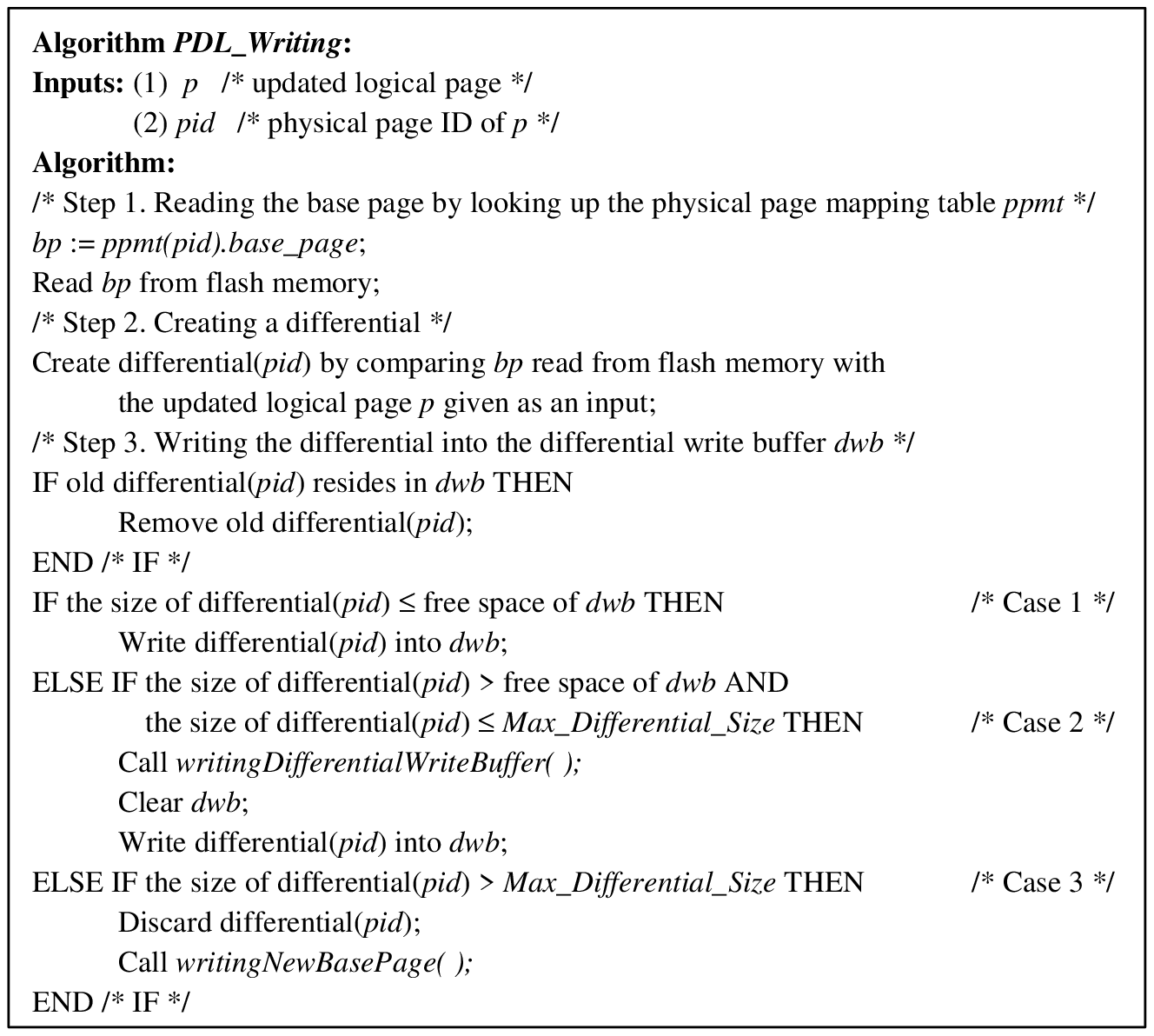, width=10.5cm}}
  \vspace*{-0.2cm}
  \caption{Writing a logical page into flash memory in PDL.}
\label{fig:4_differentiallogging_writing}
\end{figure}

\begin{figure}[h!]
  \vspace*{-0.5cm} 
  \vspace*{0.50cm}
  \centerline{\psfig{file=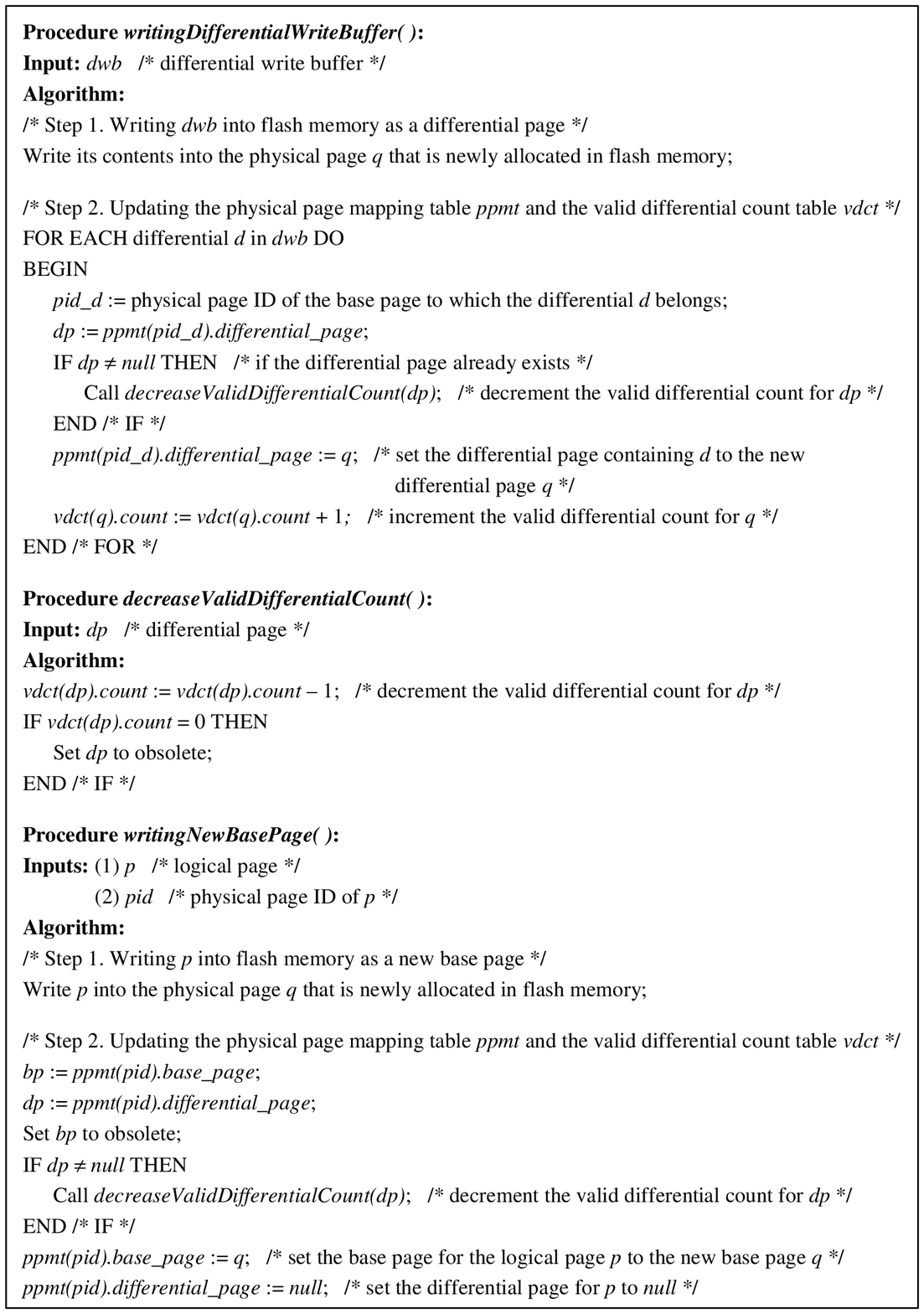, width=13.4cm}}
  \vspace*{-0.2cm}
  \caption{The procedures for the PDL\_Writing algorithm in Figure~\ref{fig:4_differentiallogging_writing}.}
\label{fig:4_differentiallogging_writing_procedure}
\end{figure}

Figure~\ref{fig:4_differentiallogging_writing} shows the algorithm
{\it PDL\_Writing}. The inputs to the algorithm are the logical page
{\it p} and its physical page ID {\it pid}. The algorithm consists
of the following three steps. In Step\,1, we read base\_page({\it
pid}) from flash memory. In Step\,2, we create differential({\it
pid}) by comparing base\_page({\it pid}) with {\it p} given as an
input. In Step\,3, we write differential({\it pid}) into the
differential write buffer. If old differential({\it pid}) resides in
the buffer, we first remove the old one, and then, write the new
one. Here, there are three cases according to the size of
differential({\it pid}). First, when the size of differential({\it
pid}) is equal to or smaller than the free space of the
buffer\,(Case\,1), we just write differential({\it pid}) into the
buffer. Second, when it is larger than the free space of the buffer
but is equal to or smaller than {\it
Max\_Differential\_Size}\,\footnote{\vspace*{-0.2cm} In
Section~\ref{chap:4_1}, for ease of exposition, we have explained
PDL on the assumption that {\it Max\_Differential\_Size} = the size
of one physical page. \vspace*{-0.2cm} However, in practice, we can
adjust it according to the workload. We will show the performance
while varying {\it Max\_Differential\_Size} later in the experiment
section\,(Section~\ref{chap:5}).}(Case\,2), we execute the procedure
{\it writingDifferentialWriteBuffer(\,)} in
Figure~\ref{fig:4_differentiallogging_writing_procedure}, clear the
buffer, and then, write differential({\it pid}) into the buffer.
Here, {\it Max\_Differential\_Size} is defined as the the maximum
size of differentials to be stored in differential pages. The
procedure {\it writingDifferentialWriteBuffer(\,)} consists of the
following two steps. In Step\,1, we write the buffer's contents into
the differential page $q$ that is newly allocated in flash memory.
In Step\,2, we update the physical page mapping table $ppmt$ and the
valid differential count table $vdct$. For each differential $d$ in
the buffer, we decrement the count for the old differential page
$dp$ in $vdct$ by executing the procedure {\it
decreaseValidDifferentialCount(\,)}. Here, if the count becomes 0,
we set the differential page to obsolete\,\footnote{\vspace*{-0.2cm}
For the spare area in a page, a write operation that changes a set
of bits from 1 to 0 can be repeatedly performed up to four times
without an erase operation\,\cite{GT05a}.} and make it available for
garbage collection. We then set differential\_page({\it pid\_d}) in
$ppmt$ to the new differential page $q$ and increment the count for
$q$ in $vdct$. Here, $pid\_d$ is the physical page ID of the base
page to which the differential {\it d} belongs. Third, when it is
larger than {\it Max\_Differential\_Size}\,(Case\,3), we discard
differential({\it pid}) and execute the procedure {\it
writingNewBasePage(\,)} in
Figure~\ref{fig:4_differentiallogging_writing_procedure}. The
procedure consists of the following two steps. In Step\,1, we write
the logical page $p$ itself into the base page $q$ that is newly
allocated in flash memory. In Step\,2, we update $ppmt$ and $vdct$.
We set the old base page $bp$ to obsolete making it available for
garbage collection. We then decrement the count for the old
differential page $dp$ in $vdct$ by executing the procedure
decreaseValidDifferentialCount(\,) and set base\_page({\it pid}) and
differential\_page({\it pid}) in $ppmt$ to $q$ and $null$,
respectively.
Figure~\ref{fig:4_differentiallogging_writing_procedure} shows the
procedures for the PDL\_Writing algorithm.

\begin{figure}[h!]
  \vspace*{0.50cm}
  \centerline{\psfig{file=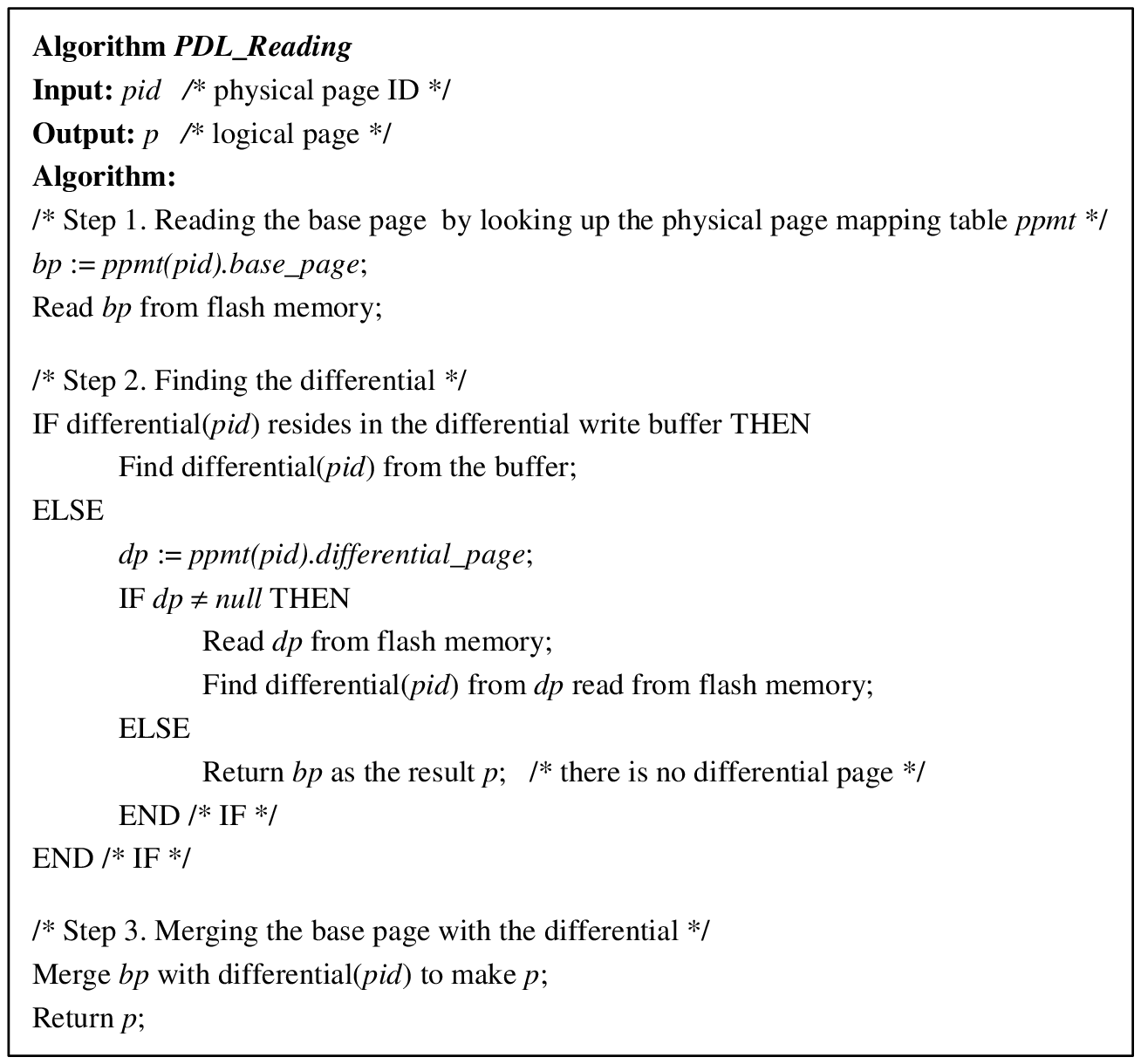, width=10.3cm}}
  \vspace*{-0.2cm}
  \caption{Recreating a logical page from flash memory in PDL.}
\label{fig:4_differentiallogging_reading}
\end{figure}

Figure~\ref{fig:4_differentiallogging_reading} shows the algorithm
{\it PDL\_Reading}. The input to PDL\_Reading is the physical page
ID {\it pid} of the logical page to read. The algorithm consists of
the following three steps. In Step\,1, we read base\_page({\it pid})
from flash memory. In Step\,2, we find differential({\it pid}) of
the base\_page({\it pid}). Here, there are two cases depending on
the place where the differential({\it pid}) resides. First, when the
differential({\it pid}) resides in the differential write buffer,
i.e., when the buffer has not been yet written out to flash memory,
we find it from the buffer. Second, when we cannot find it from the
buffer, we read differential\_page({\it pid}) from flash memory,
finding differential({\it pid}) from it. In Step\,3, we recreate a
logical page {\it p} by merging base\_page({\it pid}) read in
Step\,1 with differential({\it pid}) found in Step\,2.

\vspace*{-0.1cm}
\subsection{Discussions}
\label{chap:4_4}
\vspace*{-0.1cm}

PDL has the following four advantages. (1) As compared with
the page-based methods, it has good write performance, i.e., it
requires fewer write operations, when we need to reflect an updated
logical page into flash memory. This is due to the
writing-difference-only principle.
(2) As compared with the log-based methods, it has good write
performance when a logical page is updated multiple times. This is
due to the at-most-one-page writing principle.
(3) As compared with the log-based methods, it has good read
performance when recreating a logical page from flash memory. This
is due to the at-most-two-page reading principle.
(4) Moreover, it allows existing disk-based DBMSs to be reused
without modification as flash-based DBMSs because it is
DBMS-independent.

\begin{figure}[h!]
  \vspace*{0.50cm}
  \centerline{\psfig{file=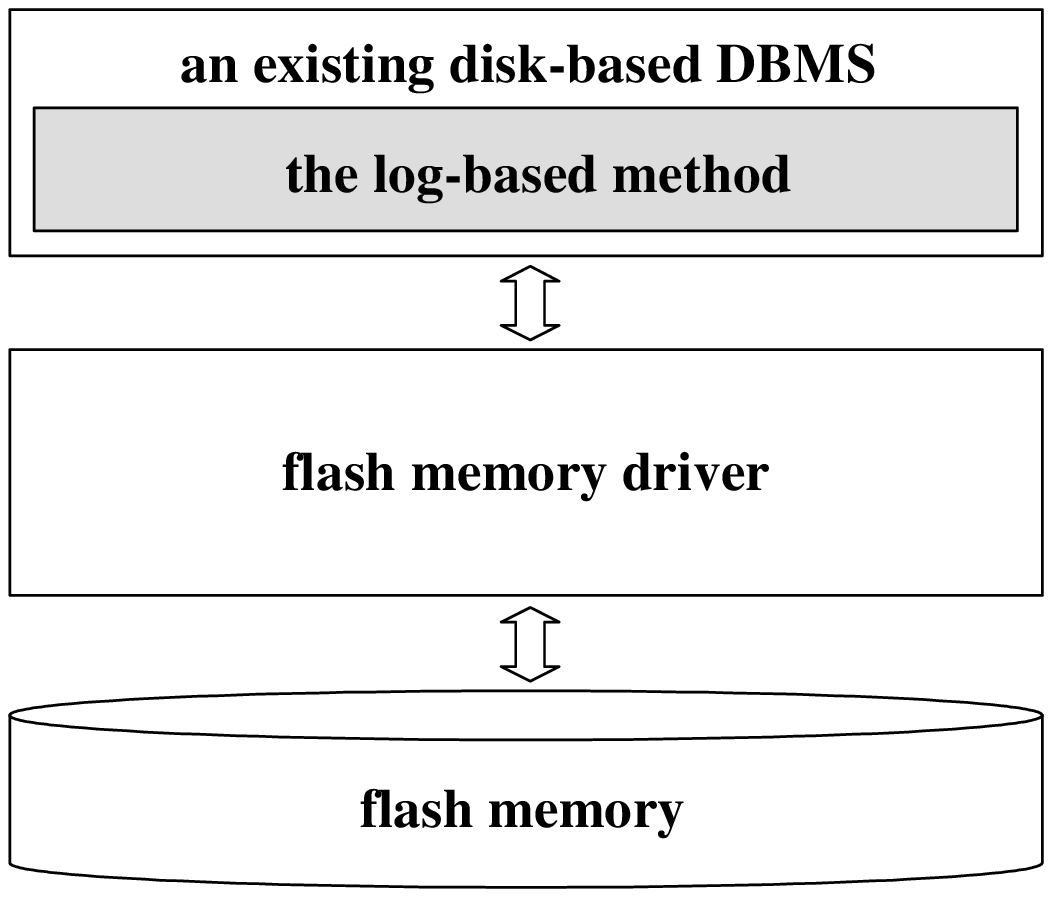, width=6cm} \hspace*{1.5cm}
              \psfig{file=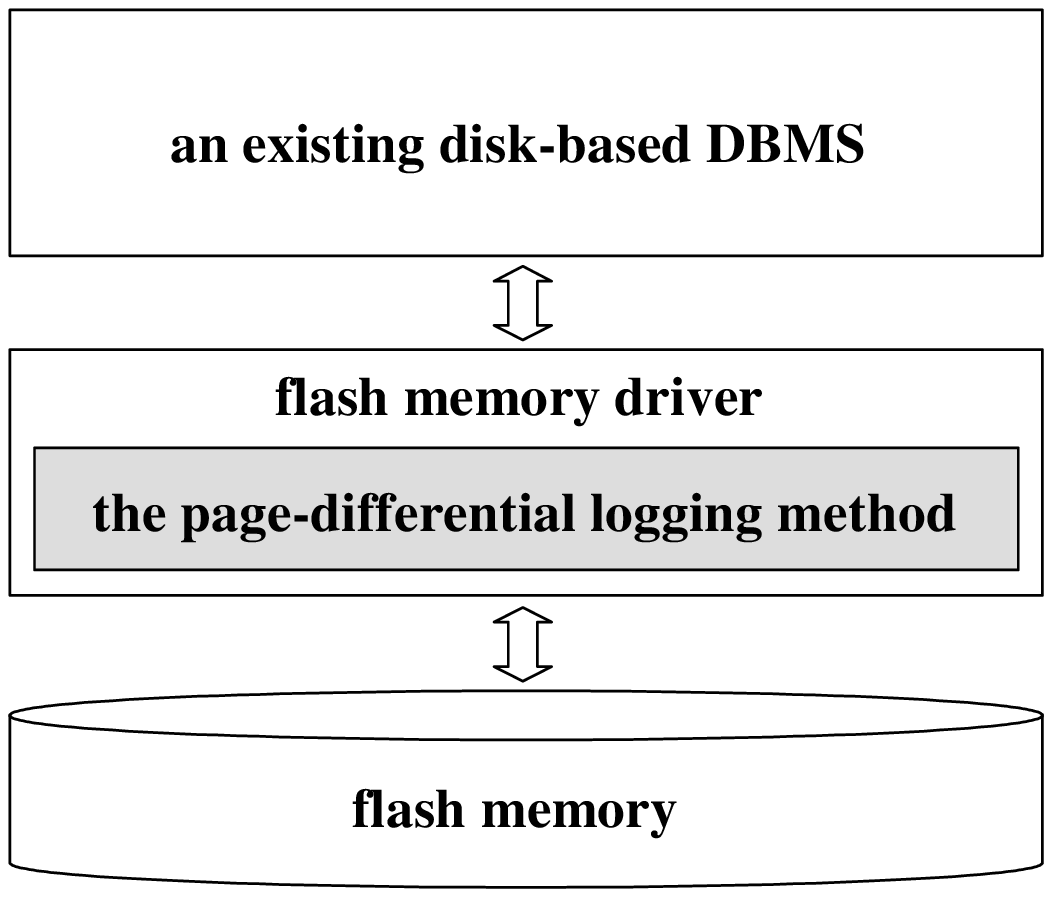, width=6cm}}
  \centerline{\hspace*{1.0cm} (a) The log-based methods. \hspace*{2.2cm}
              (b) page-differential logging.}
  \vspace*{-0.2cm}
  \caption{The DBMS architecture that uses flash memory as a secondary storage.}
\label{fig:4_differentiallogging_architecture}
\end{figure}

Figure~\ref{fig:4_differentiallogging_architecture} shows the DBMS
architecture that uses flash memory as a secondary storage. The
log-based methods need to modify the storage management module of
the DBMS so as to write the update log whenever the page is updated
as shown in
Figure~\ref{fig:4_differentiallogging_architecture}\,(a). On the
other hand, PDL does not need to modify the DBMS but to modify only
the flash memory driver\,\footnote{This flash memory driver
corresponds to the FTL shown in Figure~\ref{fig:3_ftl}.} because it
computes the differential by comparing the whole updated logical
page with its base page. Thus, it can be implemented inside the
flash memory driver as shown in
Figure~\ref{fig:4_differentiallogging_architecture}\,(b) without
affecting the storage manager of the existing DBMS.

PDL, however, has the following minor drawbacks. First, when
recreating a logical page from flash memory,
PDL has to read one more page than page-based methods do. However,
this drawback is relatively minor because the speed of read
operation is much faster than that of write or erase operations.
Furthermore, if a database is used for read-only access, {\it PDL
reads only one physical page} just like page-based methods since a
differential page does not exist\,(i.e., the base page has not been
updated). Thus, in this case, the read performance of PDL is as good
as that of the page-based methods. Second, the data size written
into flash memory in PDL could be larger than that in log-based
methods. It is because the differential contains all the difference
between an updated logical page and its base page, while the update
log in the log-based methods contains only the difference between an
updated logical page and its immediate previous version. However, in
spite of this drawback, PDL improves the overall performance
significantly because the advantages outweigh these drawbacks. We
will show the performance advantages later in the experiment
section\,(Section~\ref{chap:5}).
Table~\ref{tbl:4_differentiallogging_comparison} summarizes the
differences between PDL and the log-based ones.

\vspace*{0.50cm}
\renewcommand{\baselinestretch}{1.10}
\begin{table}
\begin{center}
\caption{Comparison of PDL with log-based and page-based ones.}
\vspace*{0.3cm}
\begin{tabular} {|c|c|c|c|}
\hline
                                    & PDL                               & log-based methods          & page-based methods            \\
\hline \hline
data to be written                  &                                   & an update log              & the whole page                \\
into flash memory                   & differential                      & (changed parts only)       & (changed and                  \\
                                    &                                   &                            & unchanged parts)              \\
\hline
time for writing data               & only when a logical page          & whenever a page is         &                               \\
into {\it the write buffer}         & needs to be reflected             & updated                    & no write buffer               \\
                                    & into flash memory                 &                            &                               \\
\hline
time for writing data               & \multicolumn{2}{c|}{}                                          & when a page needs             \\
into {\it flash memory}             & \multicolumn{2}{c|}{when the write buffer is full}             & to be reflected               \\
                                    & \multicolumn{2}{c|}{}                                          & into flash memory             \\
\hline
number of physical                  & maximum two pages                 &                            &                               \\
pages to read when                  & ($1 \leq n \leq 2$)               & multiple pages             & one page                      \\
recreating a logical page           &                                   &                            &                               \\
\hline
architecture                        & loosely-coupled                   & tightly-coupled            & loosely-coupled               \\
                                    & (DBMS-independent)                & (DBMS-dependent)           & (DBMS-independent)            \\
\hline
\end{tabular}
\label{tbl:4_differentiallogging_comparison}
\end{center}
\vspace*{0.4cm} 
\end{table}
\renewcommand{\baselinestretch}{2.0}

\vspace*{-0.1cm}
\subsection{Crash Recovery}
\label{chap:4_5}
\vspace*{-0.1cm}

A storage device with a cache normally supports a {\it
write-through} command that flushes the data written into the cache
immediately out to the device. When the write-through command is
called, PDL flushes the differential write buffer out into flash
memory. In flash memory, the page writing is guaranteed to be atomic
at the chip level\,\cite{KKCJNM01}.

When a system failure occurs, we lose the physical page mapping
table and the valid differential count table in memory. However, by
one scan through physical pages in flash memory, we can reconstruct
those tables.
Here, the tables are recovered to the state in which data were
reflected into flash memory by the write-through call or by flushing
the differential write buffer. That is, the data retained in the
write buffer only but not written out to flash memory are not
recovered in the tables. This is analogous to the situation where
data retained only in the file buffer but not written out to disk in
a disk file system are not recovered after a system failure. Thus,
when persistency of data is required, a write-through call must be
used.

If a system failure occurs when a base page\,(or the differential
write buffer) is written into flash memory, but the old base
page\,(or the differential page that does not contain any valid
differential) has not yet been set to obsolete in
Figure~\ref{fig:4_differentiallogging_writing}, the new base
page\,(or differential page) and the old base page\,(or differential
page) might co-exist in flash memory. Thus, to identify the most
up-to-date base page\,(or differential page), we use the creation
time stamp stored in a base page and in each differential in a
differential page as in Chang et al.\,\cite{CK05}.

\begin{figure}[h!]
  \vspace*{0.50cm}
  \centerline{\psfig{file=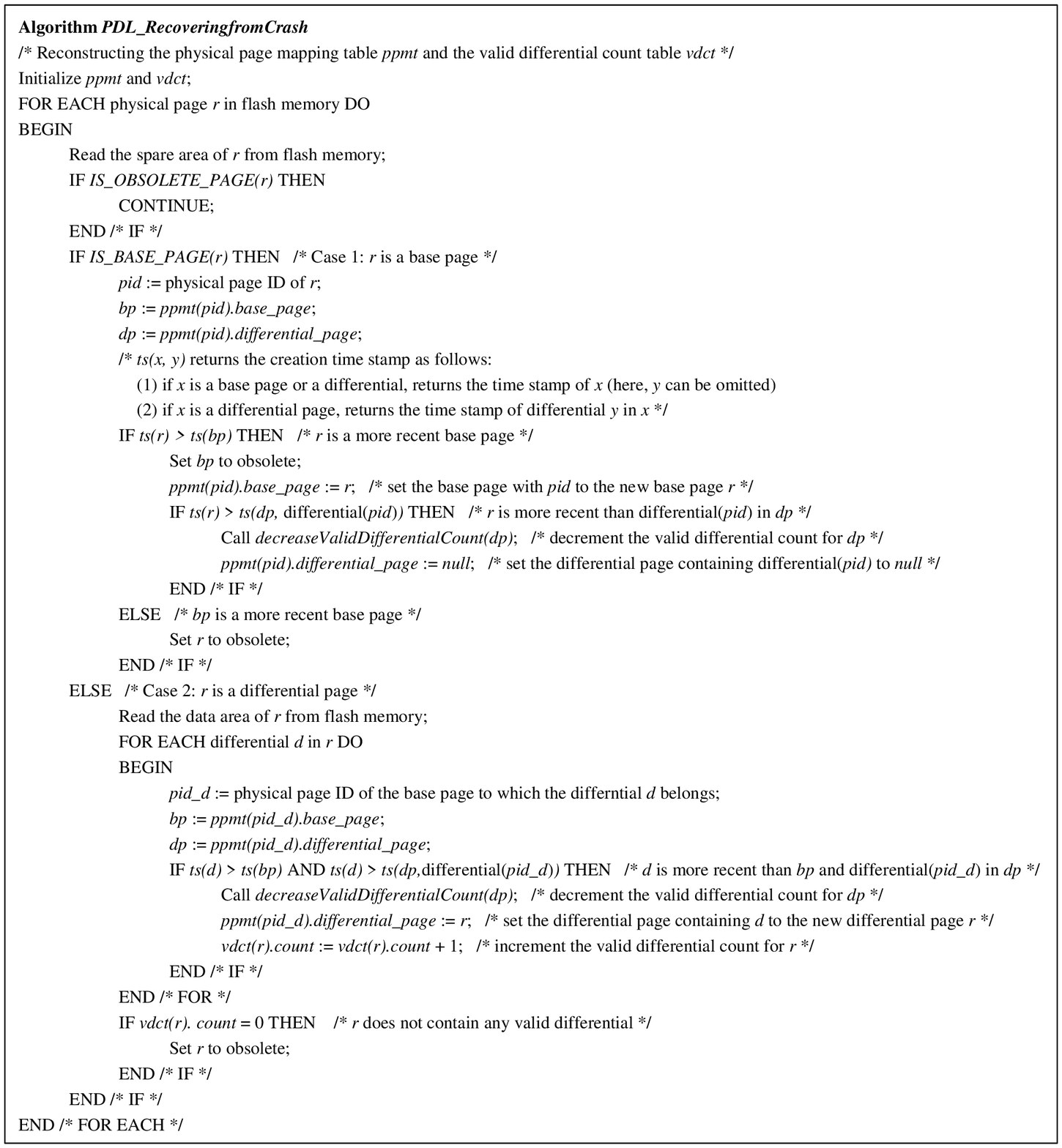, width=16.4cm}}
  \vspace*{-0.2cm}
  \caption{The algorithm for reconstructing the physical page mapping table and the valid differential count table upon system failure.}
\label{fig:4_differentiallogging_reconstructing}
\end{figure}

Figure~\ref{fig:4_differentiallogging_reconstructing} shows the
algorithm for reconstructing the physical page mapping table $ppmt$
and the valid differential count table $vdct$. For every physical
page $r$ in flash memory, we read the spare area of $r$ and update
$ppmt$ and $vdct$ only if $r$ is not obsolete. Here, there are two
cases according to the type of $r$. First, when $r$ is a base
page\,(Case 1), we check whether ts({\it r}) is more recent than
ts({\it bp}), where ts({\it r}) is the creation time stamp of $r$
and ts({\it bp}) is that of the base page $bp$ currently in $ppmt$.
If so, $r$ must be a more recent base page. Thus, we set
base\_page({\it pid}) to $r$ and set the old base page $bp$ to
obsolete, where $pid$ is the physical page ID of $r$. We then check
whether ts({\it r}) is more recent than ts({\it dp},
differential({\it pid})), which is the time stamp of
differential({\it pid}) in the differential page $dp$ currently in
$ppmt$. If so, the differential({\it pid}) must be obsolete since we
have a base page $r$ that is more recent. Thus, we set
differential\_page({\it pid}) to $null$ and decrement the count for
the old differential page $dp$ by executing the procedure
decreaseValidDifferentialCount(\,). If ts({\it r}) is not more
recent than ts({\it bp}), we set $r$ to obsolete. Second, when $r$
is a differential page\,(Case 2), we read the data area of $r$. For
each differential $d$ in $r$, we check whether ts({\it d}) is more
recent than both ts({\it bp}) and ts({\it dp}, differential({\it
pid\_d})), where ts({\it d}) is the time stamp of $d$, ts({\it bp})
is that of the base page $bp$ currently in $ppmt$, and ts({\it dp},
differential({\it pid\_d})) is that of differential({\it pid\_d}) in
the differential page $dp$ currently in $ppmt$. Here, $pid\_d$ is
the physical page ID of the base page to which the differential {\it
d} belongs. If so, $d$ must be a more recent differential of $bp$
than differential({\it pid\_d}) currently in $ppmt$. Thus, we set
differential\_page({\it pid\_d}) to $r$, decrement the count for the
old differential page $dp$ by executing the procedure
decreaseValidDifferentialCount(\,), and increment the count for the
new differential page $r$. If $r$ does not contain any valid
differential after processing all the differentials in $r$, we set
$r$ to obsolete.

In Figure~\ref{fig:4_differentiallogging_reconstructing}, we set two
kinds of useless pages to obsolete: (1) base pages that are not
recent but have not been set to obsolete and (2) differential pages
that do not contain valid differential but have not been set to
obsolete. These pages can occur in flash memory when a system
failure occurs if a base page\,(or the differential write buffer)
has been written into flash memory, but the old base page\,(or the
differential page that does not contain valid differentials) has not
yet been set to obsolete.

The algorithm PDL\_RecoveringfromCrash guarantees that recovery is
normally performed even when a system failure repeatedly occurs
during the process of restarting the system. The reason is that the
algorithm does not change data in the flash memory except setting
the useless pages\,(i.e., the pages that are no longer used, but
have not been set to obsolete) to obsolete. Setting useless pages to
obsolete does not affect the recovery process of reconstructing the
physical page mapping table and the valid differential count table.

Since scanning the entire flash memory of 1\,Gbytes takes
approximately 60 seconds\,(derived from
Table~\ref{tbl:2_flashmemory} in Section~\ref{chap:2}), the scan
time can be practically accommodated. To recover the physical page
mapping table without scanning all the physical pages in flash
memory, we have to log the changes in the mapping table into flash
memory. We leave this extension as a further study.

We note that we can implement the proposed PDL and recovery
techniques in a DBMS that uses flash memory to support transactional
database recovery just as we do in a DBMS built on top of an O/S
file system by using the write-through facility whenever persistency
of a write operation is required\,(e.g., when writing the
`transaction commit' log record).

%
%
\section{Performance Evaluation}
\label{chap:5}
\vspace*{-0.30cm}

\vspace*{-0.1cm}
\subsection{Experimental Data and Environment}
\label{chap:5_1}
\vspace*{-0.1cm}

We compare the data access performance of PDL proposed in this paper
with those of the page-based and log-based methods discussed in
Section~\ref{chap:3}. We use the wall clock time taken to access
data from flash memory\,(we call it {\it the I/O time}) as the
measure. Here, as the page-based method, we use the one employing
the out-place update\,(OPU) scheme with the page-level mapping
technique, which is known to have good performance even though the
method consumes memory excessively\,\cite{KKCJNM01}. We also compare
with the in-place update method\,(IPU). As the log-based method, we
use the in-page logging method\,(IPL) proposed by Lee and
Moon\,\cite{LM07}.

We use the synthetic relational data of 1\,Gbytes and update
operations for comparing data access performance of the three
methods. We define an {\it update} operation as consisting of the
following three steps: (1) reading the addressed page; (2) changing
the data in the page; and (3) writing the updated page. The reading
step\,(1) creates a logical page by reading physical pages from
flash memory, and the writing step\,(3) writes the updated logical
page as one or more physical pages into flash memory. The
experiments are designed this way to exclude the buffering effect in
the DBMS. Therefore, we can measure read, write as well as overall
performance by executing only update operations.

The I/O time is affected by $N\_updates\_till\_write$ and
$\%ChangedByOneU\_Op$. Here, $N\_updates\_till\_write$ is the number of update
operations applied to a logical page in memory from the time it is
recreated from flash memory until the time it is reflected back into
flash memory, $\%ChangedByOneU\_Op$ is the percentage of data
changed in a logical page by a single update operation. Here, the
portion of data to be changed is randomly selected.
We also compare the performance of various mixes of read-only and
update operations varying the percentage of the update
operations\,($\%UpdateOps$). Besides, we measure the performance as
we vary the performance parameters of flash memory\,(i.e., the I/O
times for read and write operations in
Table~\ref{tbl:2_flashmemory}). We also compare the longevity of
flash memory. Finally, we perform the TPC-C benchmark\,\cite{TPC02}
as a real workload. Table~\ref{tbl:5_experiment} summarizes the
experiments and parameters.

\vspace*{-0.50cm} 
\vspace*{0.50cm}
\renewcommand{\baselinestretch}{1.10}
\begin{table}
\begin{center}
\caption{Experiments and parameters.}
\vspace*{0.3cm}
\begin{tabular} {|c|c|c|c|}
\hline
\multicolumn{2}{|c|}{Experiments} & \multicolumn{2}{c|}{Parameters} \\
\hline \hline
Exp. 1 & Read, write, and overall time per                  & $\%ChangedByOneU\_Op$        & 2                \\ \cline{3-4}
       & update operation                                   & $N\_updates\_till\_write$    & 1                \\
\hline
Exp. 2 & Overall time per update operation                  & $\%ChangedByOneU\_Op$        & 2                \\ \cline{3-4}
       & as $N\_updates\_till\_write$ is varied             & $N\_updates\_till\_write$    & 1 $\sim$ 8       \\
\hline
Exp. 3 & Overall time per update operation                  & $\%ChangedByOneU\_Op$        & 0.1 $\sim$ 100   \\ \cline{3-4}
       & as $\%ChangedByOneU\_Op$ is varied                 & $N\_updates\_till\_write$    & 1, 5             \\
\hline
       & Overall time per operation for the mixes           & $\%ChangedByOneU\_Op$        & 2                \\ \cline{3-4}
Exp. 4 & of read-only and update operations                 & $N\_updates\_till\_write$    & 1, 5             \\ \cline{3-4}
       & as $\%UpdateOps$ is varied                         & $\%UpdateOps$                & 0 $\sim$ 100     \\
\hline
       &                                                    & $\%ChangedByOneU\_Op$        & 2                \\ \cline{3-4}
Exp. 5 & Overall time per update operation as               & $N\_updates\_till\_write$    & 1                \\ \cline{3-4}
       & the parameters of flash memory are varied          & $T_{read}$                   & 10 $\sim$ 1500   \\ \cline{3-4}
       &                                                    & $T_{write}$                  & 500, 1000        \\
\hline
Exp. 6 & Number of erase operations per update              & $\%ChangedByOneU\_Op$        & 2                \\ \cline{3-4}
       & operation as $N\_updates\_till\_write$ is varied   & $N\_updates\_till\_write$    & 1 $\sim$ 8       \\
\hline
Exp. 7 & I/O time per transaction for TPC-C data            &                              & 1 $\sim$ 100\,Mbytes\\
       & as the DBMS buffer size is varied                  & DBMS buffer size             & (0.1 $\sim$ 10\,\% of\\
       &                                                    &                              & database size)   \\
\hline
\end{tabular}
\label{tbl:5_experiment}
\end{center}
\end{table}
\renewcommand{\baselinestretch}{2.0}

In each experiment, garbage collection is invoked whenever there is
no more free page in flash memory\,\footnote{In IPL, garbage
collection is invoked during the process of merging.}. Here, the
cost\,(time) of garbage collection is amortized into that of the
write operation because garbage collection is incurred by the accumulated
effect of write operations. We repeatedly execute experiments so
that garbage collection is invoked for each block at least ten times
on the average after loading the database in order to make the
database to reach a steady state.

~\newline 

For the experiments, we have implemented an emulator of a 2-Gbyte
flash memory chip using the parameters shown in
Table~\ref{tbl:2_flashmemory}\,\footnote{\vspace*{-0.2cm} For each
operation, the emulator returns the required time in the flash
memory, which is specified in Table~\ref{tbl:2_flashmemory}, while
writing and reading the data to and from the disk. \vspace*{-0.2cm}
The data are in exactly the same format in disk as would be stored
in flash memory. Thus, access time using the emulator must be
identical to that using the real flash memory.}.
We also have implemented the four methods: PDL\,($x$), OPU, IPU, and
IPL\,($y$)\,\footnote{We set the size of log buffer for each logical
page to the size of a logical page\,$\times \frac{1}{16}$ as was
used by Lee and Moon\,\cite{LM07}.}\,\footnote{\vspace*{-0.2cm} We
do not use wear-leveling in this paper, but the same wear-leveling
techniques can be applied to these methods. We use the same garbage
collection method suggested by Woodhouse\,\cite{W01}} for PDL and
OPU. Here, $x$ is {\it Max\_Differential\_Size} (defined in
Section~\ref{chap:4_3} in p.~17), and $y$ is the amount of log pages
in each block.
We used the Odysseus ORDBMS\,\cite{WLLKH05,WLKLL07} as the storage
system. Here, PDL, OPU, and IPU are implemented outside the DBMS,
and IPL inside the DBMS. We conducted all experiments on a Pentium 4
3.0\,GHz Linux PC with 2\,Gbytes of main memory. We set the size of
a logical page to be 2\,Kbytes, which is the size of a physical page
in flash memory. We also test the case with a logical page of
8\,Kbytes as was done by Lee and Moon\,\cite{LM07}.

\vspace*{-0.1cm}
\subsection{Results of the Experiments}
\label{chap:5_2}
\vspace*{-0.1cm}

\begin{figure}[hp]
  \centerline{\hspace*{1.5cm} \psfig{file=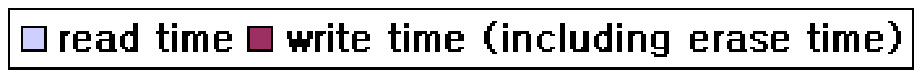, width=7cm}}
  \vspace*{0.3cm}
  \centerline{\psfig{file=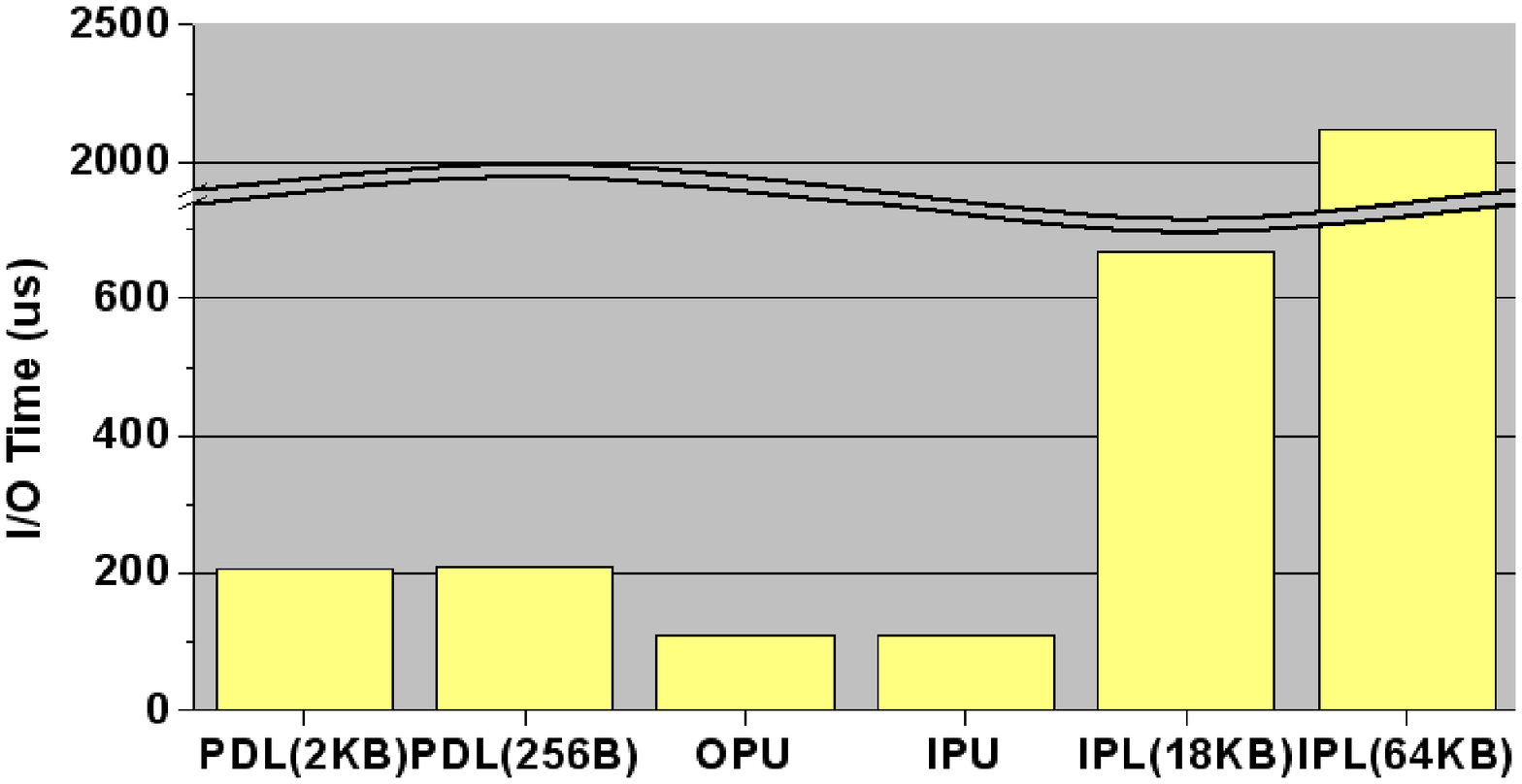, width=11.0cm}}
  \vspace*{-0.3cm}
  \centerline{(a) The I/O time of the reading step.}
  \vspace*{0.5cm}
  \centerline{\psfig{file=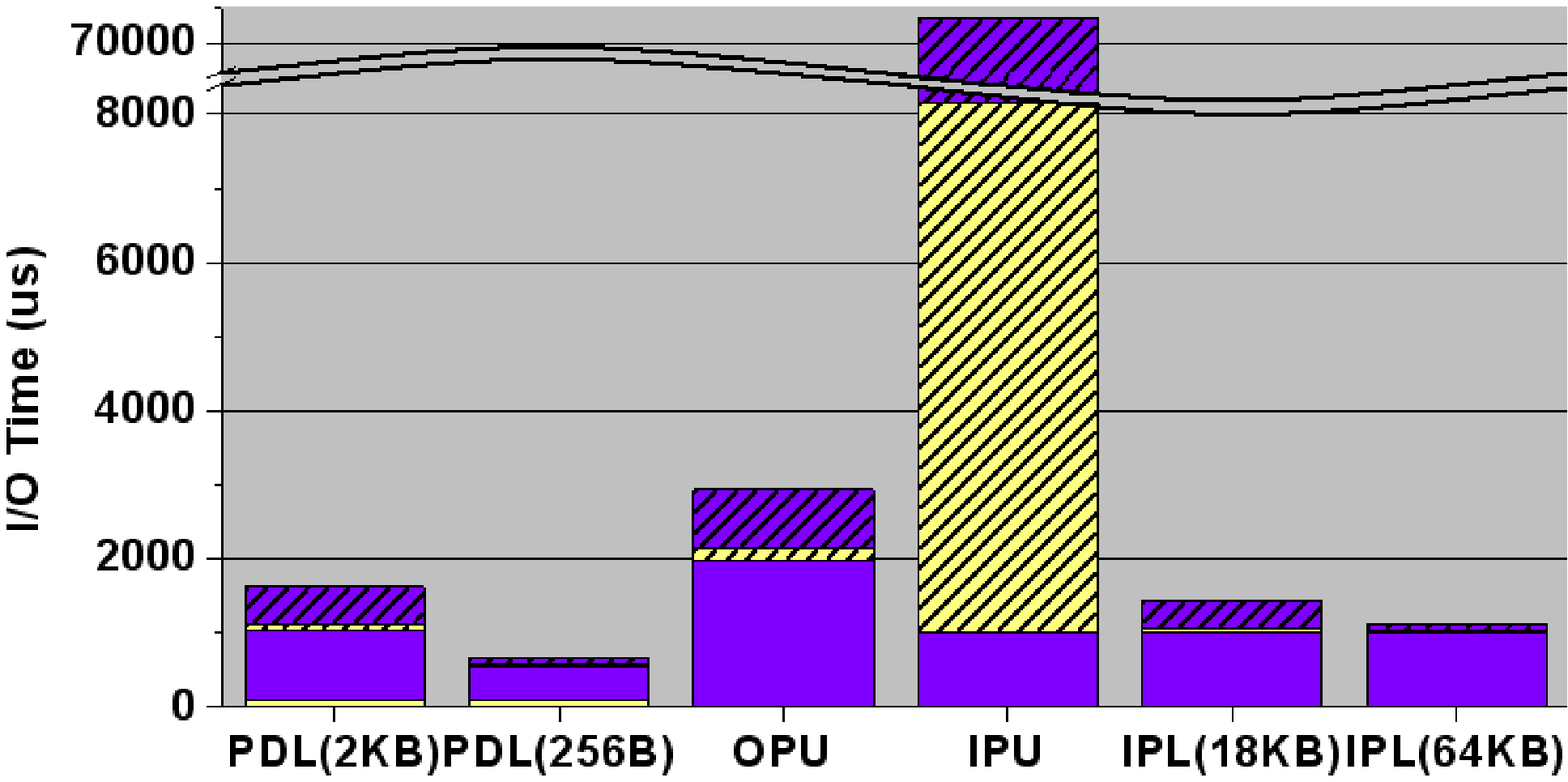, width=11.0cm}}
  \vspace*{-0.3cm}
  \centerline{(b) The I/O time of the writing step. Slashed parts indicate the}
  \vspace*{-0.3cm}
  \centerline{time for garbage collection. Lighter areas represent read time.}
  \vspace*{0.5cm}
  \centerline{\psfig{file=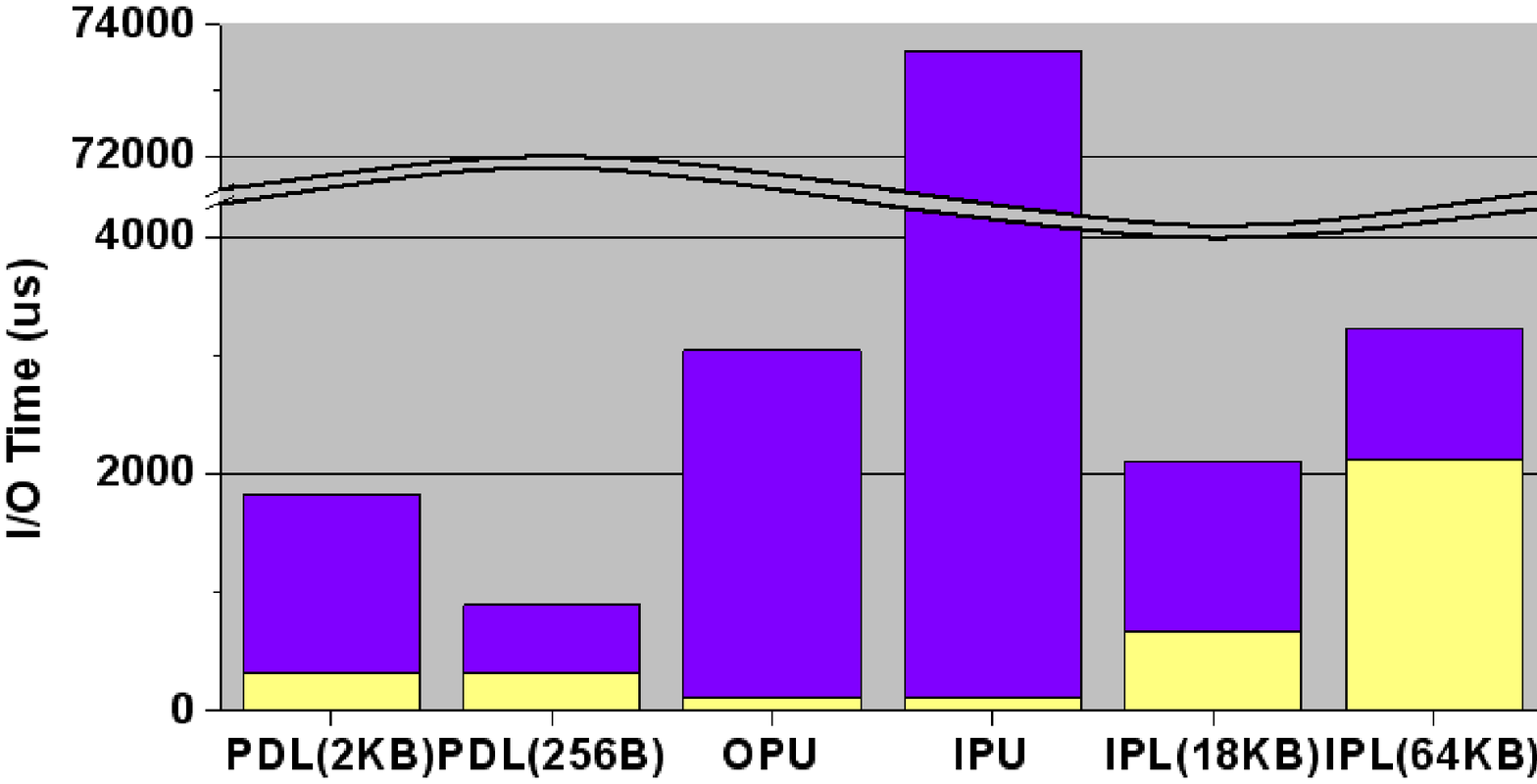, width=11.0cm}}
  \vspace*{-0.3cm}
  \centerline{(c) The overall time per update operation including read and write times in (a) and (b).}
  \vspace*{-0.2cm}
  \caption{The read, write, and overall time per update operation ($N\_updates\_till\_write = 1$, $\%ChangedByOneU\_Op = 2$, database size = 1\,Gbytes, $T_{read} = 110\,\mu s, T_{write} = 1010\,\mu s$).}
  \label{fig:5_experiment1}
\end{figure}

\noindent \textbf{Experiment~1:} \\
Figure~\ref{fig:5_experiment1} shows the read, write, and overall
time per update operation for the six methods: IPL\,(18KB),
IPL\,(64KB), PDL\,(2KB), PDL\,(256B), OPU, and IPU. For IPL\,($y$),
we have varied $y$ from 8\,Kbytes to 64\,Kbytes. Among them, we
select IPL\,(18KB) and IPL\,(64KB) because they have the best and
worst overall time for update operations, respectively. For PDL, we
select PDL\,(2KB) and PDL\,(256B) because the amounts of
differential pages in them are similar to those of log pages in
IPL\,(64KB) and IPL\,(18KB), respectively. Specifically, IPL\,(64KB)
and PDL\,(2KB) use 50\,\% of flash memory for storing
log/differential pages. IPL\,(18KB) and PDL\,(256B) use 14.1\,\% and
11.1\,\% of flash memory, respectively.

Figure~\ref{fig:5_experiment1}\,(a) shows that the I/O time of the
reading step per update operation is in the following order:
IPL\,(64KB), IPL\,(18KB), PDL\,(2KB)\,/\,PDL(256B), and OPU\,/\,IPU.
This result is consistent with what was discussed in
Sections~\ref{chap:3} and \ref{chap:4}. OPU and IPU require one read
operation. PDL requires at most twice as many read operations. IPL
requires multiple read operations. We note that, when we perform
read-only operations, we can also achieve the same result as is
shown in Figure~\ref{fig:5_experiment1}\,(a).

Figure~\ref{fig:5_experiment1}\,(b) shows that the I/O time of the
writing step is in the following order: IPU, OPU, PDL\,(2KB),
IPL\,(18KB), IPL\,(64KB), and PDL\,(256B). Here, the slashed area
indicates the I/O time for garbage collection. The result is also
consistent with the discussions in Sections~\ref{chap:3} and
\ref{chap:4}. For an update operation, OPU requires two write
operations: one for writing the updated page into flash memory and
another for setting the original page to obsolete. However, IPL
requires only one write operation for writing the log buffer into
flash memory. PDL\,(2KB) requires two write operations approximately
for every two update operations: one for writing the differential
write buffer into flash memory and another for setting one\,(on the
average) differential page to obsolete\,\footnote{When the count of
valid differentials in $vdct$ becomes 0, we set the differential
page to obsolete.} because the size of a differential is
approximately half a page on the average\,\footnote{\vspace*{-0.2cm}
Since the size of a differential changes from 0 to 1 page size and
back to 0\,(Case 3 in
Figure~\ref{fig:4_differentiallogging_writing}) as updating a
logical page is repeated, the size of a differential in a steady
state is approximately half a page on the average.}. Thus,
PDL\,(2KB) requires approximately one write operation for an update
operation on the average. PDL\,(256B) requires a less number of
write operations than PDL\,(2KB) does since the differential write
buffer is filled less frequently. But, PDL additionally requires one
read operation for reading the base page in from flash memory in
order to create the differential. Here, each method includes a
certain amount of read cost, which is incurred by garbage collection
and amortized into the write cost. We note that PDL\,(256B)
outperforms the other methods due to less frequent writing of the
differential write buffer.

\begin{figure}[h!]
  \vspace*{0.50cm}
  \centerline{\hspace*{1.2cm} \psfig{file=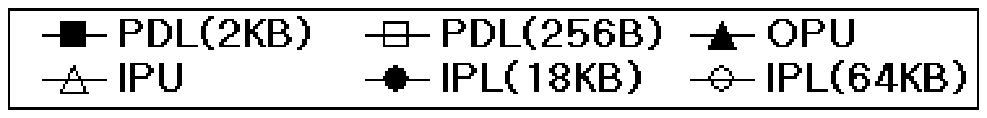, width=6.0cm}
              \hspace*{1.9cm} \psfig{file=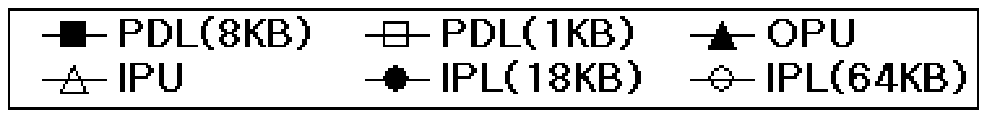, width=6.0cm}}
  \vspace*{0.3cm}
  \centerline{\psfig{file=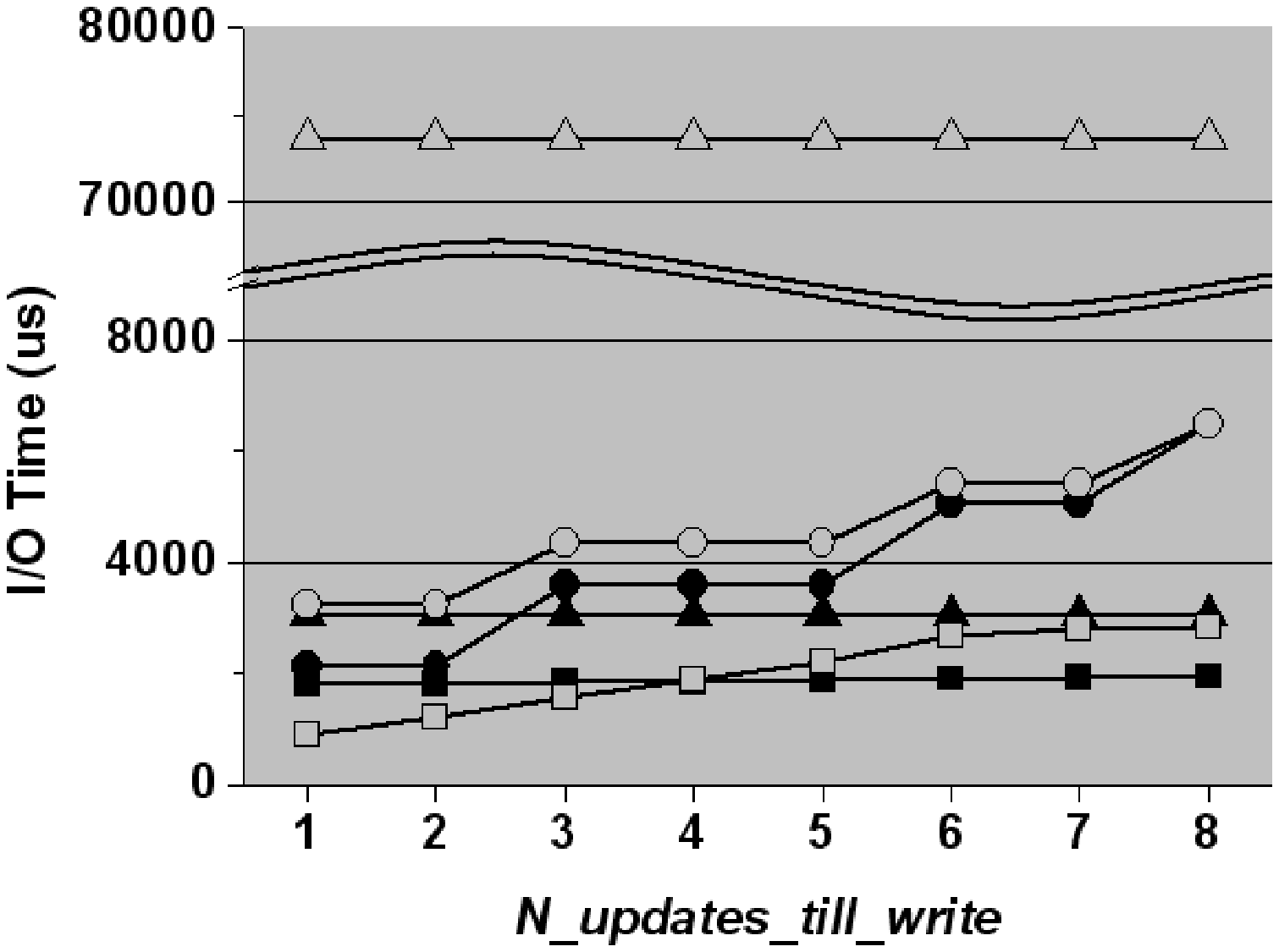, width=7.5cm}
              \hspace*{0.5cm}
              \psfig{file=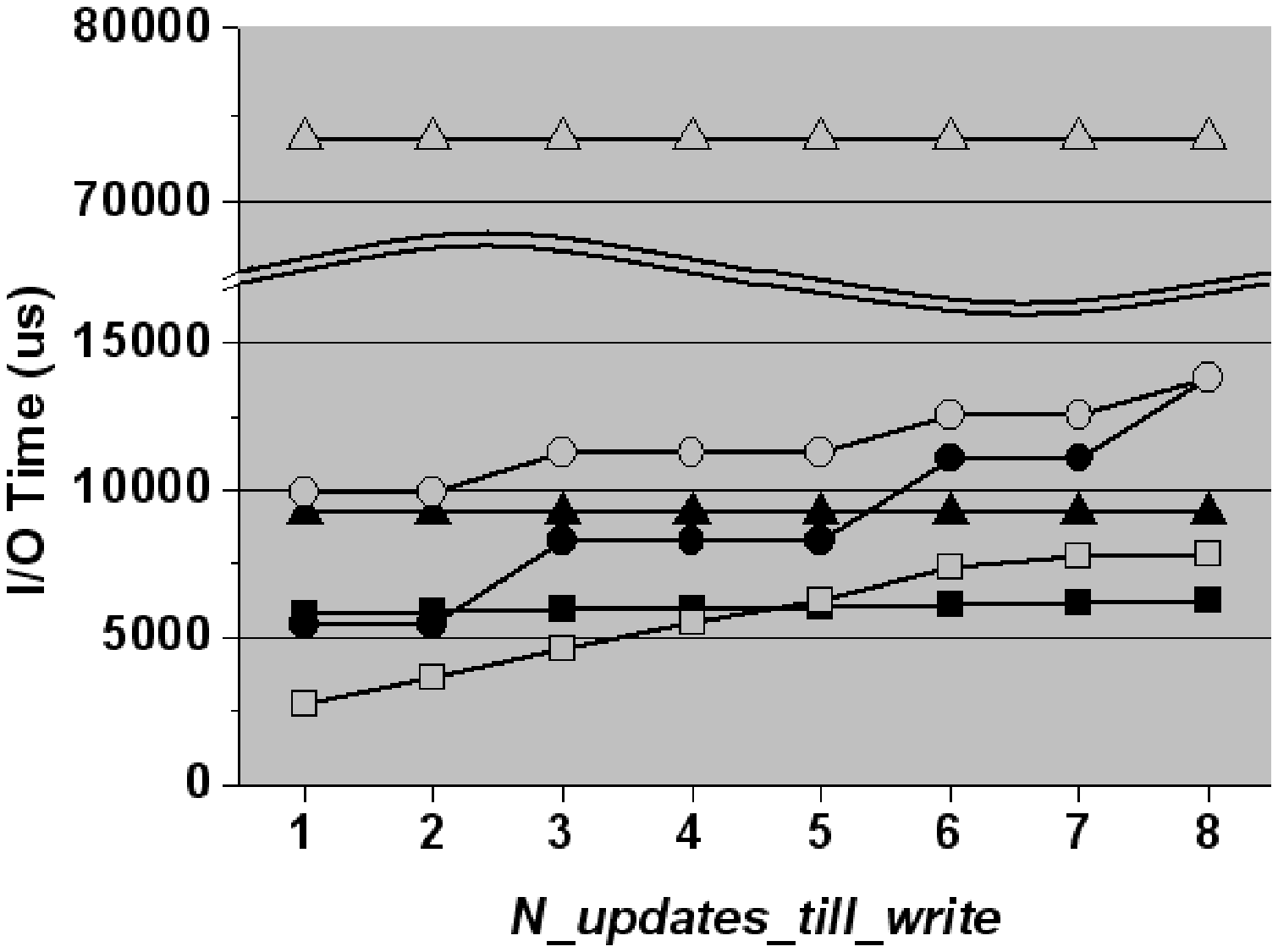, width=7.5cm}}
  \vspace*{-0.3cm}
  \leftline{\hspace*{1.5cm} (a) size of a logical page = 2\,Kbytes. \hspace*{2.5cm} (b) size of a logical page = 8\,Kbytes.}
  \vspace*{-0.2cm}
  \caption{The overall time per update operation as $N\_updates\_till\_write$ is varied ($\%ChangedByOneU\_Op = 2$).}
  \label{fig:5_experiment2}
\end{figure}

Figure~\ref{fig:5_experiment1}\,(c) shows the overall time per
update operation combining the I/O times shown in
Figures~\ref{fig:5_experiment1}\,(a) and (b). PDL\,(256B) has good
read and write performance as shown in
Figures~\ref{fig:5_experiment1}\,(a) and (b), and thus, has the best
overall time for an update operation.\,(This corresponds to
Figure~\ref{fig:5_experiment2}\,(a) when $N\_updates\_till\_write =
1$.)

~\newline \noindent \textbf{Experiment~2:} \\
Figure~\ref{fig:5_experiment2} shows the overall time per update
operation of the six methods as $N\_updates\_till\_write$ is varied.
First, the I/O time of OPU and IPU is steady regardless of the
parameter because they always write the whole page when reflecting
an updated logical page into flash memory. Next, the I/O time of IPL
increases in a stepwise manner. The reason for this behavior is that
the number of write operations for a logical page is computed as
$\lceil \frac{the~size~of~update~logs}{the~size~of~log~buffer}
\rceil$. Here, the size of the update logs to be written increases
linearly as $N\_updates\_till\_write$ increases because IPL keeps
all the update logs of a logical page. (We note that this process of
writing is not bound by merging while the reading process is.)
Finally, the I/O time of PDL\,(2KB) increases only very slightly as
$N\_updates\_till\_write$ increases because the size of the overlap
among the changed parts becomes larger as $N\_updates\_till\_write$
increases with the total size of the difference being limited to one
page. The I/O time of PDL\,(256B) increases approximately linearly
as $N\_updates\_till\_write$ increases because the size of the
overlap is small. As $N\_updates\_till\_write$ increases, the I/O
time of PDL\,(256B) approaches that of OPU because the logical page
itself\,(rather than the differential) is written into flash memory
as the size of the differential becomes larger than {\it
Max\_Differential\_Size}\,(Case 3 in
Figure~\ref{fig:4_differentiallogging_writing}). As a result,
PDL\,(256B) outperforms OPU, IPU, and IPL.
The result when the size of a logical page is 8\,Kbytes shows a
tendency similar to that when the size of a logical page is
2\,Kbytes.

\begin{figure}[h!]
  \vspace*{0.50cm}
  \centerline{\psfig{file=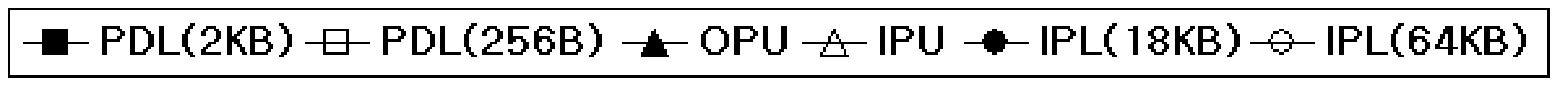, width=9.0cm}}
  \vspace*{0.3cm}
  \centerline{\psfig{file=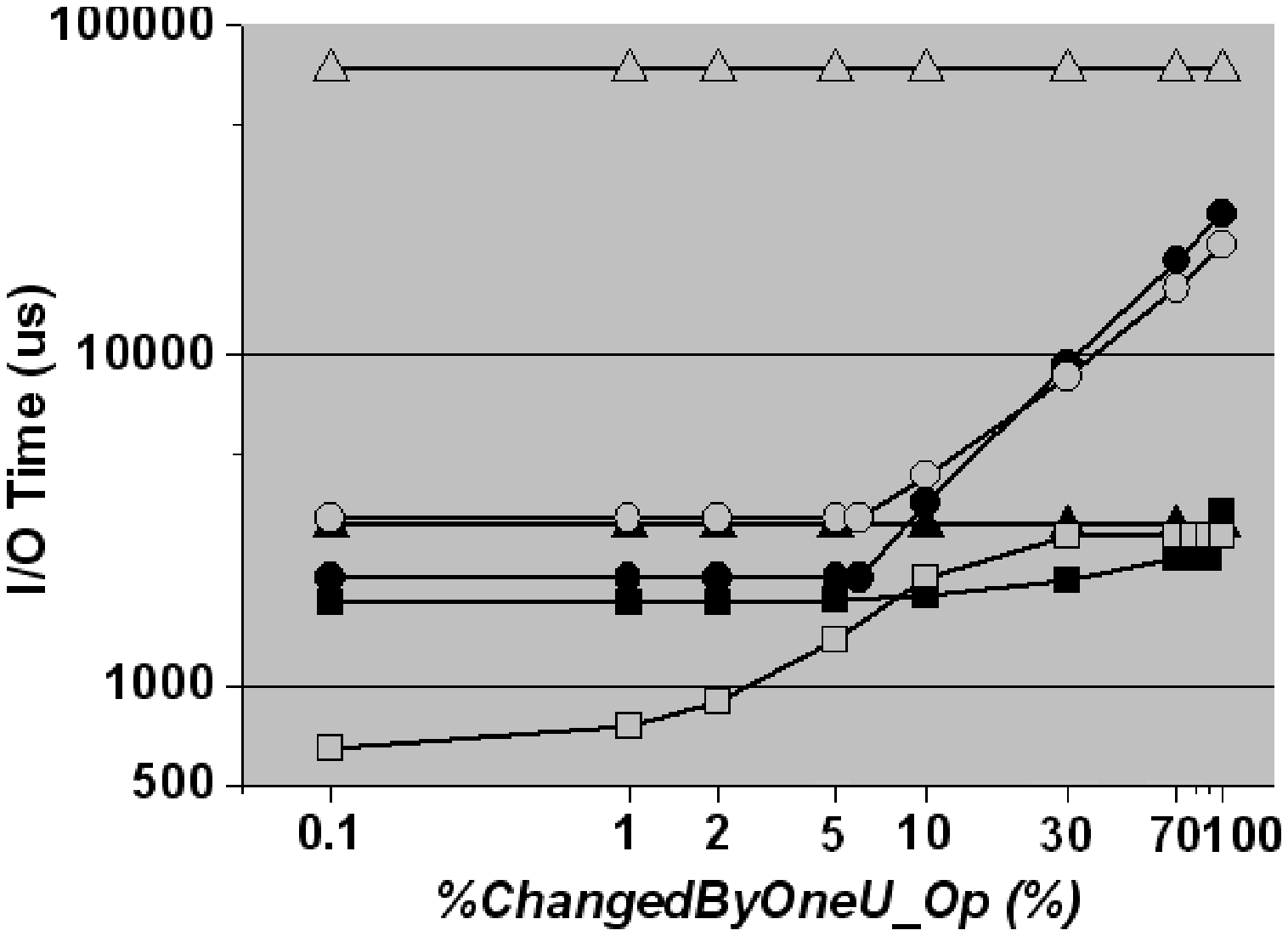, width=7.5cm}
              \hspace*{0.5cm}
              \psfig{file=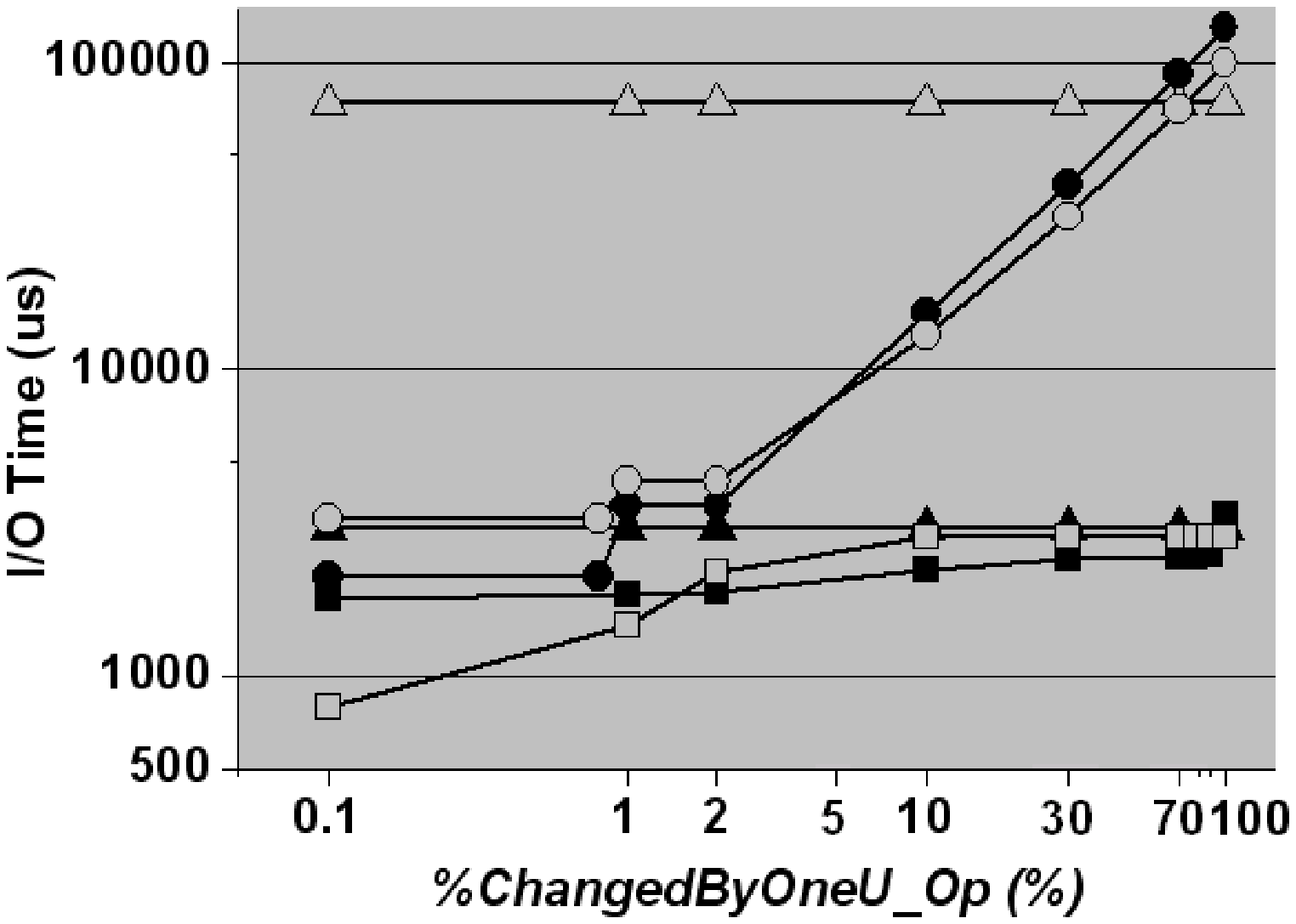, width=7.5cm}}
  \vspace*{-0.3cm}
  \leftline{\hspace*{2.0cm} (a) $N\_updates\_till\_write = 1$. \hspace*{3.5cm} (b) $N\_updates\_till\_write = 5$.}
  \vspace*{-0.2cm}
  \caption{The overall time per update operation as $\%ChangedByOneU\_Op$ is varied ($N\_updates\_till\_write = 1, 5$).}
  \label{fig:5_experiment3}
\end{figure}

~\newline \noindent \textbf{Experiment~3:} \\
Figure~\ref{fig:5_experiment3} shows the overall time per update
operation for the six methods as $\%ChangedByOneU\_Op$ is varied.
The result is consistent with what we observed in
Figure~\ref{fig:5_experiment2}.
We note that PDL\,(256B) outperforms OPU, IPU, and IPL for the same
reason as in Figure~\ref{fig:5_experiment2}. When
$\%ChangedByOneU\_Op \approx 100$, the I/O time of PDL\,(2KB) is
slightly larger than that of OPU because, while the two methods
require the same number of write operations, PDL\,(2KB) needs three
times as many read operations\,---\,for reading the base page and
the differential page when recreating a logical page from flash
memory, and then, for reading the base page again to create the
differential when reflecting the updated logical page into flash
memory.

\begin{figure}[h!]
  \vspace*{0.50cm}
  \centerline{\psfig{file=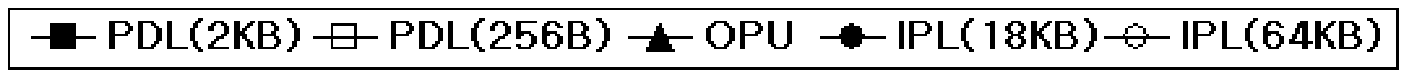, width=8.6cm}}
  \vspace*{0.3cm}
  \centerline{\psfig{file=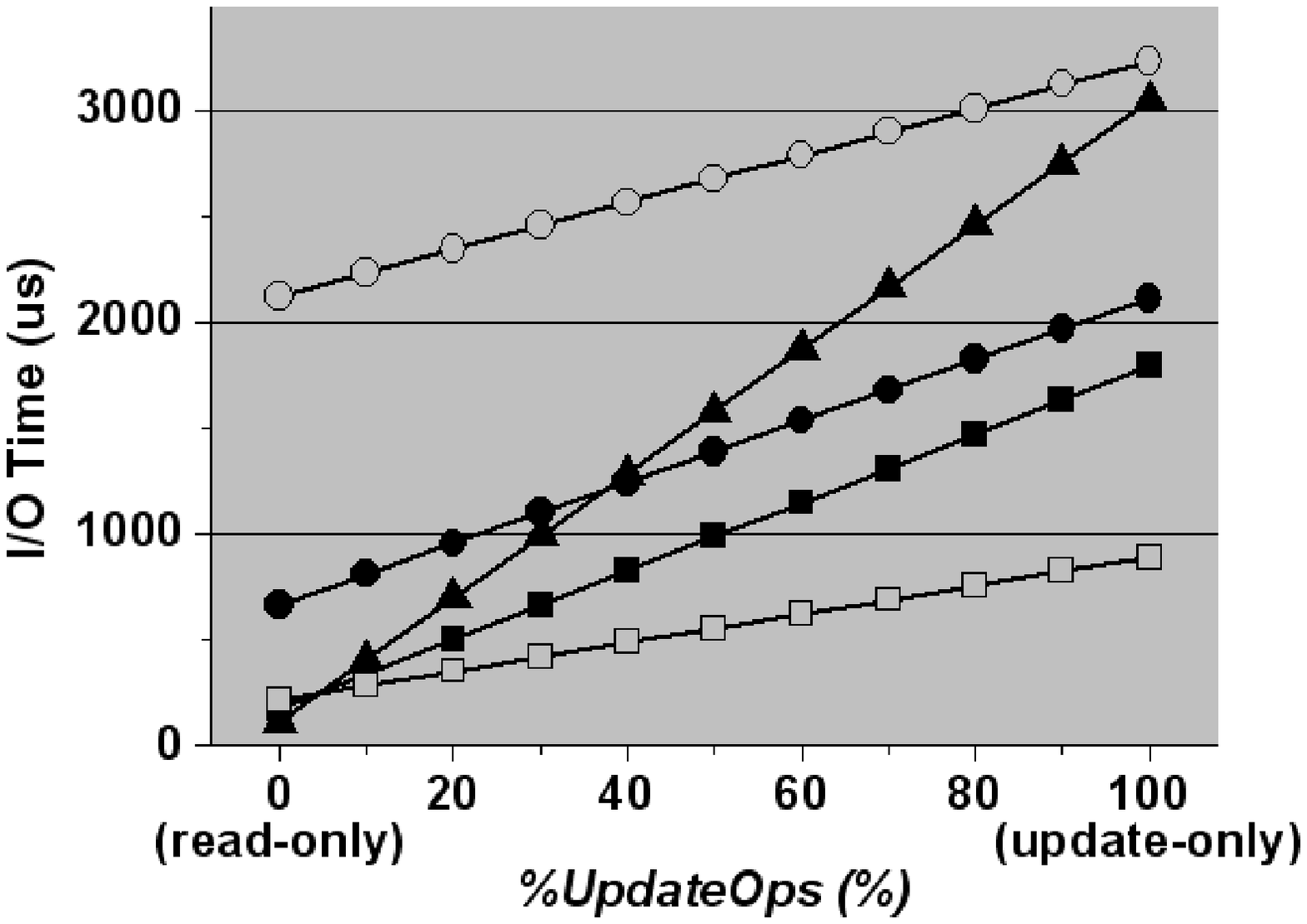, width=7.5cm}
              \hspace*{0.5cm}
              \psfig{file=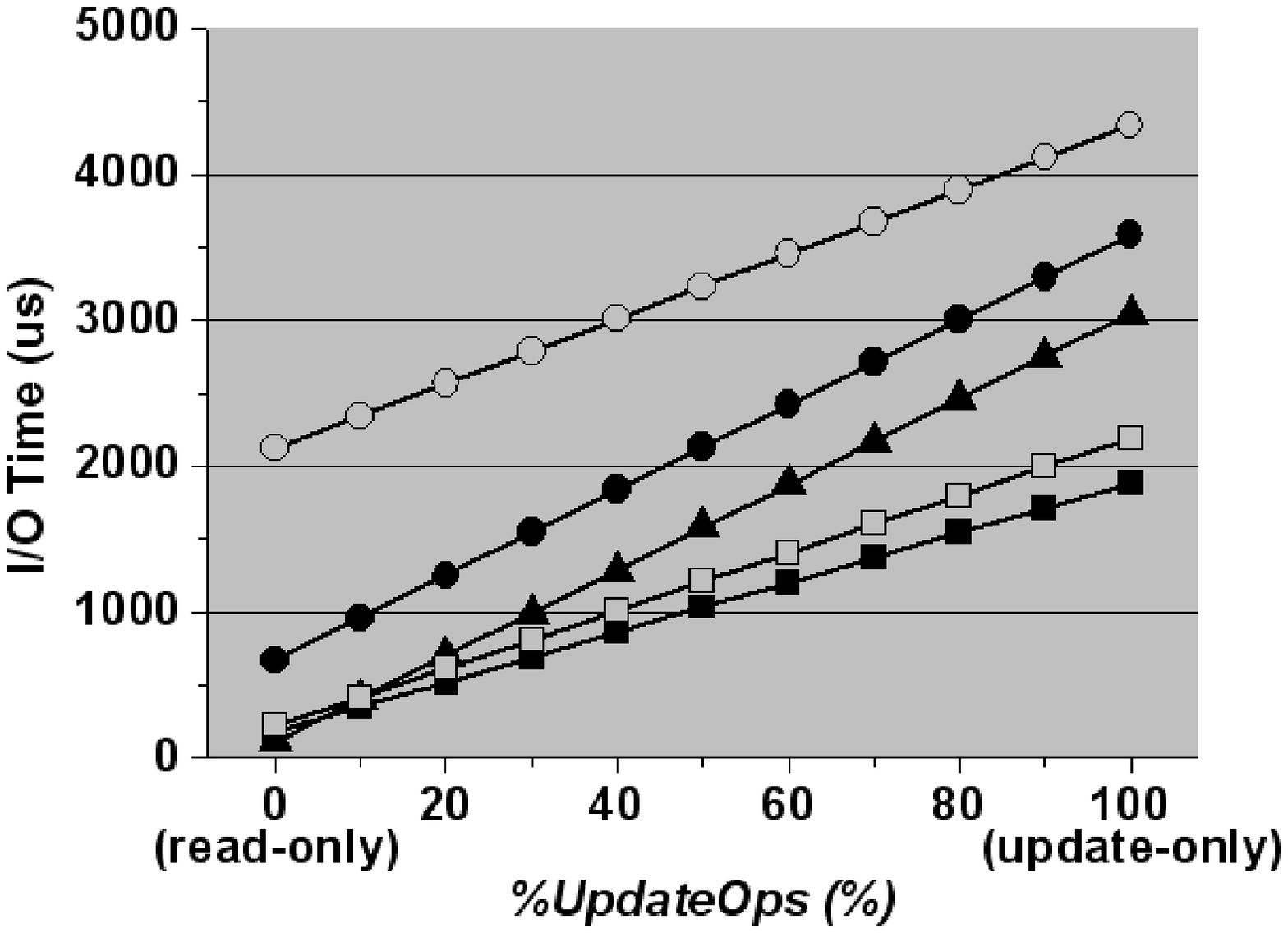, width=7.5cm}}
  \vspace*{-0.3cm}
  \leftline{\hspace*{1.75cm} (a) $N\_updates\_till\_write = 1$. \hspace*{3.5cm} (b) $N\_updates\_till\_write = 5$.}
  \vspace*{-0.2cm}
  \caption{The overall time per operation for the mixes of read-only and update operations as $\%UpdateOps$ is varied ($\%ChangedByOneU\_Op = 2$).}
  \label{fig:5_experiment4}
\end{figure}

~\newline \noindent \textbf{Experiment~4:} \\
Figure~\ref{fig:5_experiment4} shows the results of Experiment~4.
When updates are rare\,(i.e., $\%UpdateOps \approx 0$), OPU
outperforms PDL and IPL\,(see Figure~\ref{fig:5_experiment1}\,(a)).
As $\%UpdateOps$ increases, PDL becomes superior to OPU because of
its superiority in update performance\,(see
Figure~\ref{fig:5_experiment1}\,(c)). We also note that PDL always
outperforms IPL. In summary, for various mixes of read-only and
update operations, PDL\,(256B) improves performance by 0.5 $\sim$
3.4 times over OPU and by 1.6 $\sim$ 3.1 times over IPL\,(18KB) and
by 2.0 $\sim$ 9.7 times over IPL\,(64KB). We note that the case of
0.5 times over OPU is the special case where all transactions are
read-only\,(i.e., $\%UpdateOps = 0$).

\begin{figure}[h!]
  \vspace*{0.50cm}
  \centerline{\psfig{file=5_experiment4_caption1.eps, width=8.6cm}}
  \vspace*{0.3cm}
  \centerline{\psfig{file=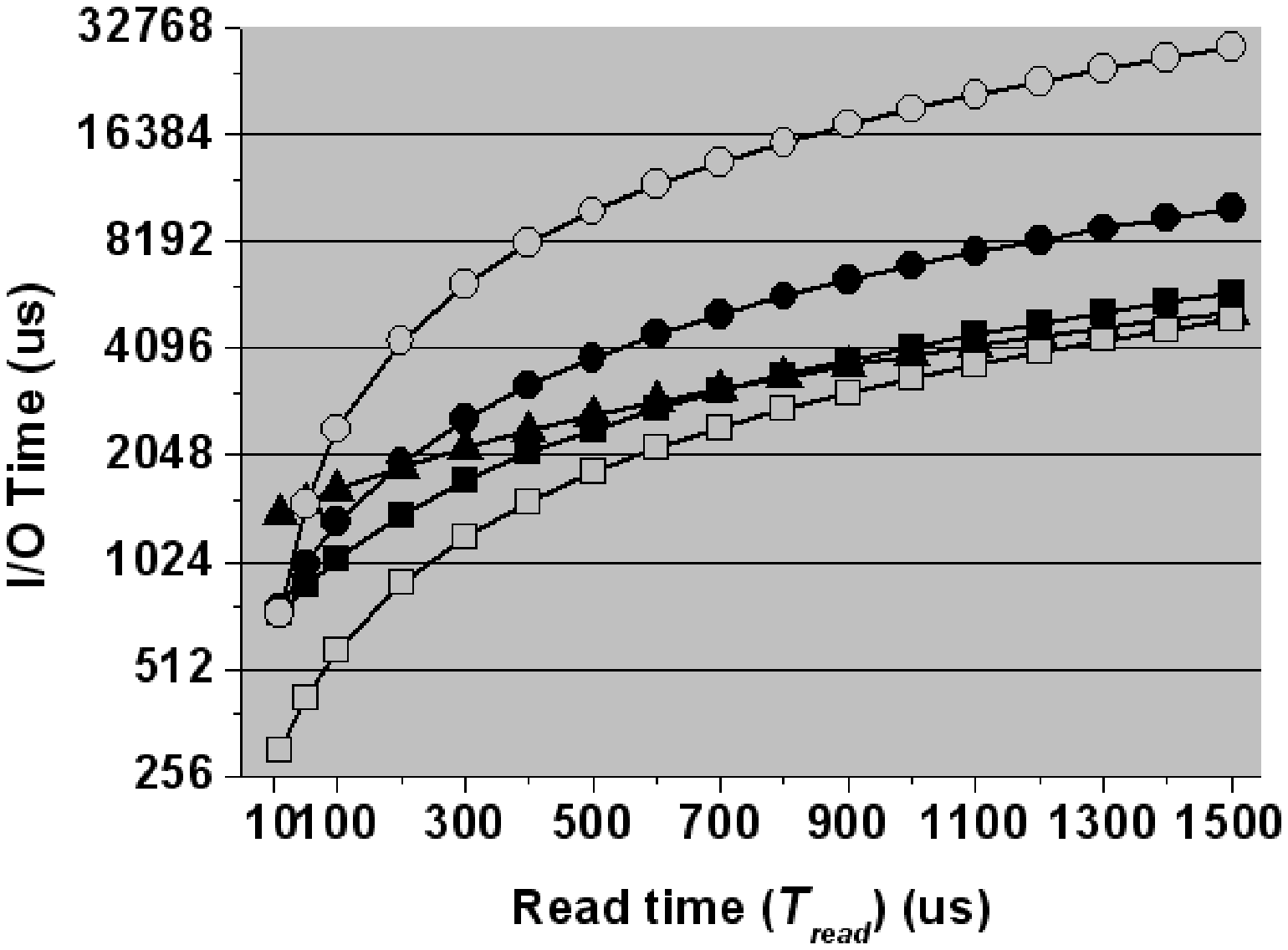, width=7.5cm}
              \hspace*{0.5cm}
              \psfig{file=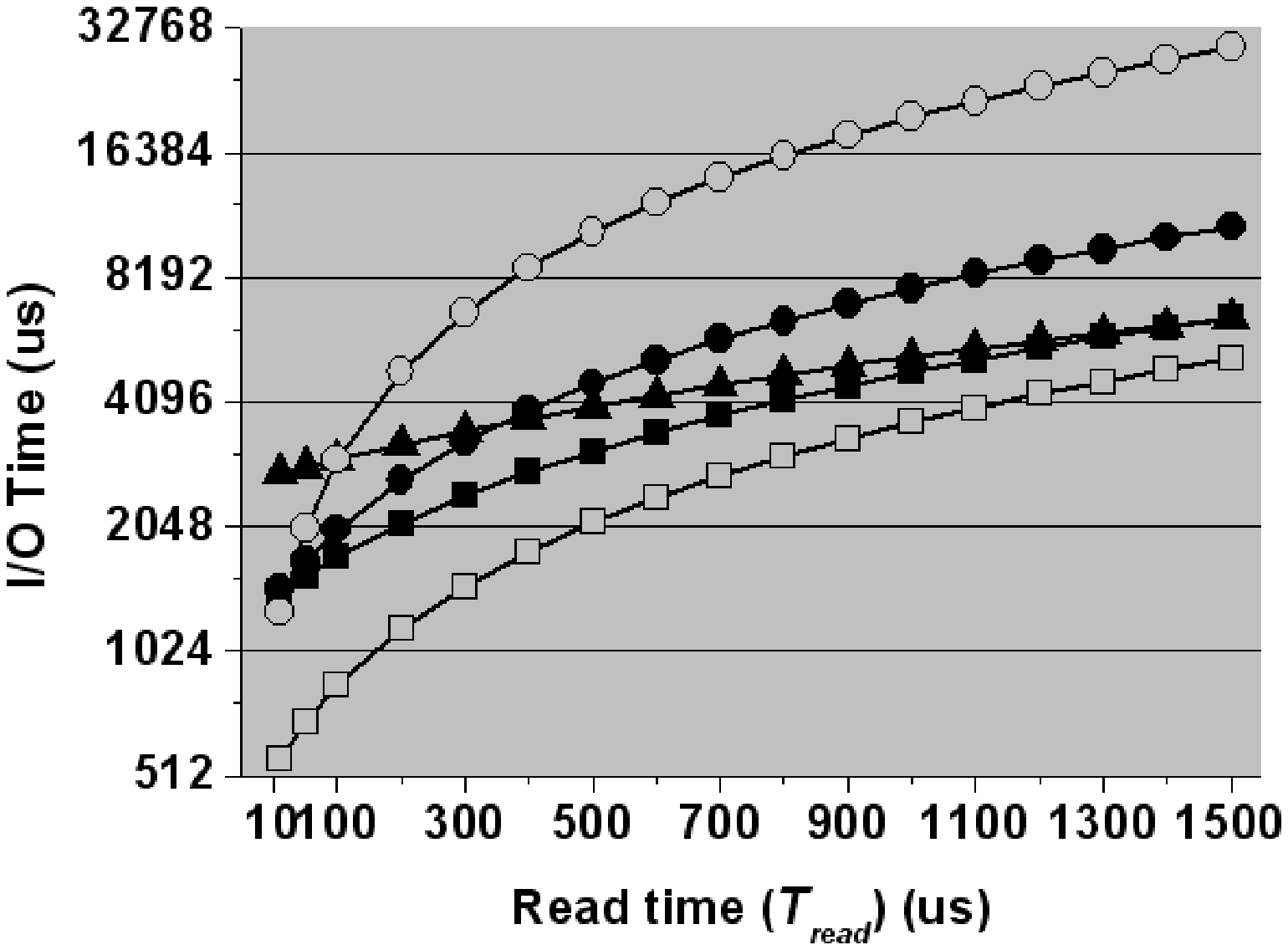, width=7.5cm}}
  \vspace*{-0.3cm}
  \leftline{\hspace*{2.5cm} (a) $T_{write}$ = $500\,\mu s$. \hspace*{5.0cm} (b) $T_{write}$ = $1000\,\mu s$.}
  \vspace*{-0.2cm}
  \caption{The overall time per update operation as the $T_{read}$ and $T_{write}$ parameters of flash memory are varied ($N\_updates\_till\_write = 1, \%ChangedByOneU\_Op = 2, T_{erase} = 1500\,\mu s$).}
  \label{fig:5_experiment5}
\end{figure}

~\newline \noindent \textbf{Experiment~5:} \\
Figure~\ref{fig:5_experiment5} shows the overall time per update
operation as the $T_{read}$ and $T_{write}$ parameters of flash
memory are varied. We observe that PDL\,(256B) always outperforms
OPU and IPL. As the read time\,($T_{read}$) increases, OPU becomes
superior to PDL\,(2KB) or IPL. We have this result because OPU has
superiority in read performance\,(see
Figure~\ref{fig:5_experiment1}\,(a)). We note that PDL\,(256B)
outperforms OPU and IPL regardless of the $T_{read}$ and $T_{write}$
parameters of flash memory.

\begin{figure}[h!]
  \vspace*{0.50cm}
  \centerline{\hspace*{1.0cm} \psfig{file=5_experiment4_caption1.eps, width=8.6cm}}
  \vspace*{0.3cm}
  \centerline{\psfig{file=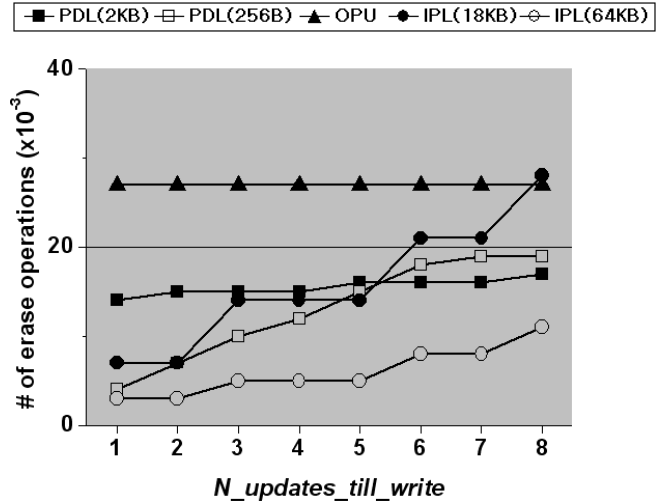, width=7.5cm}}
  \vspace*{-0.2cm}
  \caption{The number of erase operations per update operation as $N\_updates\_till\_write$ is varied ($\%ChangedByOneU\_Op = 2$).}
  \label{fig:5_experiment6}
\end{figure}

~\newline \noindent \textbf{Experiment~6:} \\
Figure~\ref{fig:5_experiment6} shows the number of erase operations
per update operation as $N\_updates\_till\_write$ is varied. We
observe that, when $N\_updates\_till\_write = 1$, the number of
erase operations per update operation is in the following order:
OPU, PDL\,(2KB), IPL\,(18KB), PDL\,(256B), and IPL\,(64KB). Thus,
IPL\,(64KB) has the best longevity among the five methods. But, it
has poor performance for the mixes of read-only and update
operations as shown in Figure~\ref{fig:5_experiment4}. PDL\,(256B)
has good longevity next to IPL\,(64KB). Besides, it has
significantly good performance for the mixes of read-only and update
operations.

\begin{figure}[h!]
  \vspace*{0.50cm}
  \centerline{\hspace*{1.0cm} \psfig{file=5_experiment4_caption1.eps, width=8.6cm}}
  \vspace*{0.3cm}
  \centerline{\psfig{file=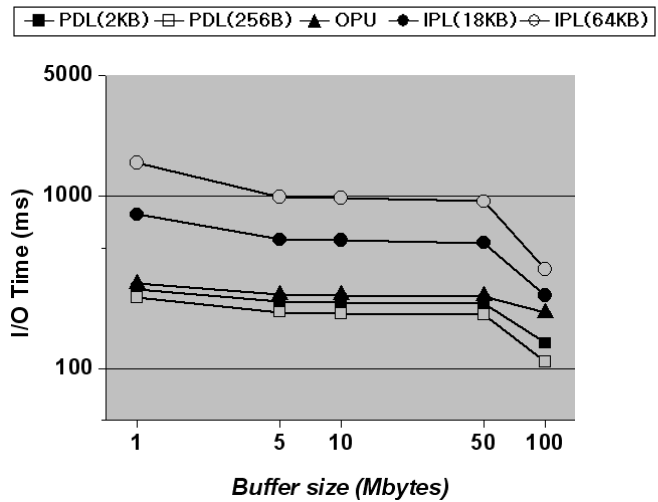, width=7.5cm}}
  \vspace*{-0.2cm}
  \caption{TPC-C benchmark: I/O time per transaction as the DBMS buffer size is varied (database size = 1\,Gbytes).}
  \label{fig:5_experiment7}
\end{figure}

~\newline \noindent \textbf{Experiment~7:} \\
Figure~\ref{fig:5_experiment7} shows the results of the TPC-C
benchmark. We observe that the I/O time is in the following order:
IPL\,(64KB), IPL\,(18KB), OPU, PDL\,(2KB), and PDL\,(256B).
The result shows that PDL outperforms other methods in real
workloads as well.

%
%
\section{Conclusions}
\label{chap:6}
\vspace*{-0.30cm}

We have proposed a novel approach for storing data called {\it
page-differential logging} for flash-based storage systems. We have
defined the notion of the differential and presented the algorithms
for reading and writing pages into flash memory using the
differential.

We have identified three design principles: writing-difference-only,
at-most-one-page writing, and at-most-two-page reading. These
principles guarantee good performance for both read and write
operations. We have shown that our method conforms to these
principles.

Page-differential logging is DBMS-independent, i.e., it allows
existing disk-based DBMSs to be reused as flash-based DBMSs just by
modifying the flash memory driver. In addition, it improves the
longevity of flash memory by reducing the number of erase operations
compared with existing page-based methods.

We have performed extensive experiments to compare the performance
of page-differential logging with existing page-update methods.
Through these experiments, we have shown that the performance of our
method is superior to those of page-based and log-based
methods\,---\,except when all transactions are read-only on already
updated pages.
We also performed experiments as the performance figures of read and
write operations change. The results show that our method\,(in
particular, PDL\,(256B)) is always superior to other methods. Thus,
the results indicate that page-differential logging can be the
preferred technique for commercial
products\,\footnote{\vspace*{-0.2cm} Commercial SSD's offer average
write time comparable to read time by exploiting parallelism, but
individual NAND flash chips typically have asymmetric read/write
times.}. We also performed experiments to compare various methods
for the longevity of flash memory. The results show that our
method\,(in particular, PDL\,(256B)) improves the longevity of flash
memory compared with OPU and IPL\,(18KB). Finally, we performed the
TPC-C benchmark as the DBMS buffer size is varied. The results show
that our method\,(in particular, PDL\,(256B)) outperforms other
methods by 1.2 $\sim$ 6.1 times. This shows effectiveness of our
method under real workloads.

Currently, we are implementing page-differential logging on a flash
memory embedded board.
Such an augmented board is to be incorporated to our Odysseus
DBMS\,\cite{WLLKH05,WLKLL07}. The resulting system will facilitate
various flash-memory-dependent optimizations in various components
of the DBMS such as the indexes, buffer, sort module, and query
optimizer. We also note that, due to its DBMS-independent nature,
page-differential logging can be employed by the manufacturer in the
FTL of commercial SSD's. We leave these issues as future work.

%
%
\section{Acknowledgement}
\vspace*{-0.3cm}

This research was partially supported by the National Research Lab
Program through the National Research Foundation of Korea\,(NRF)
funded by the Ministry of Education, Science, and
Technology\,(No.~2009-0083120). Page-differential logging has been
introduced as a part of Yi-Reun Kim's Ph.D. dissertation\,\cite{K09}
and patented in Korea\,\cite{WK09} in Nov. 2009.

%
%

\end{document}